\documentclass[10pt,letterpaper]{article}
\usepackage[top=0.85in,left=1in,footskip=0.75in]{geometry}

\usepackage{amsmath,amssymb}

\usepackage{changepage}

\usepackage[utf8x]{inputenc}

\usepackage{textcomp,marvosym}

\usepackage{cite}

\usepackage{nameref,hyperref}

\usepackage[right]{lineno}

\usepackage{microtype}
\DisableLigatures[f]{encoding = *, family = * }

\usepackage[table]{xcolor}

\usepackage{array}

\newcolumntype{+}{!{\vrule width 2pt}}

\newlength\savedwidth

\raggedright
\usepackage[aboveskip=1pt,labelfont=bf,labelsep=period,justification=raggedright,singlelinecheck=off]{caption}

\makeatletter
\renewcommand{\@biblabel}[1]{\quad#1.}
\makeatother

\usepackage{lastpage,fancyhdr,graphicx}
\usepackage{textgreek}
\usepackage{subfig}
\usepackage{booktabs}
\usepackage{epstopdf}
\fancyheadoffset[L]{2.25in}
\fancyfootoffset[L]{2.25in}
\begin{document}
\vspace*{0.2in}

\begin{flushleft}
{\Large
\textbf\newline{Learning diffusion coefficients, kinetic parameters, and the number of underlying states from a multi-state diffusion process: robustness results and application to PDK1/PKC\textalpha\, dynamics} 
}
\newline
\\
L.R. Baker\textsuperscript{1*},
M.T. Gordon\textsuperscript{2},
B.P. Ziemba\textsuperscript{2},
V. Gershuny\textsuperscript{3},
J.J. Falke\textsuperscript{2},
D.M. Bortz\textsuperscript{1*},
\\
\bigskip
\textbf{1} Department of Applied Mathematics, University of Colorado, Boulder, CO, USA
\\
\textbf{2} Department of Biochemistry, University of Colorado, Boulder, CO, USA
\\
\textbf{3} Division of Applied Regulatory Science, Office of Clinical Pharmacology, Office of Translational Sciences, Center for Drug Evaluation and Research, US Food and Drug Administration, Silver Spring, MD, USA
\\
\bigskip

%
%





* lewis.r.baker@colorado.edu, david.bortz@colorado.edu

\end{flushleft}
\section*{Abstract}
Systems driven by Brownian motion are ubiquitous. A prevailing challenge
is inferring, from data, the diffusion and kinetic parameters that describe these stochastic
processes. In this work, we investigate a multi-state diffusion process
that arises in the context of single particle tracking (SPT), wherein
the motion of a particle is governed by a discrete set of diffusive
states, and the tendency of the particle to switch between these states
is modeled as a random process. We consider two models for this behavior:
a mixture model and a hidden Markov model (HMM). For both, we adopt
a Bayesian approach to sample the distributions of the underlying
parameters and implement a Markov Chain Monte Carlo (MCMC) scheme
to compute the posterior distributions, as in \cite{coombs2009}.
The primary contribution of this work is a study of the robustness
of this method to infer parameters of a three-state HMM, and a discussion
of the challenges and degeneracies that arise from considering three
states. Finally, we investigate the problem of determining the number
of diffusive states using model selection criteria. We present results
from simulated data that demonstrate proof of concept, as well as
apply our method to experimentally measured single molecule diffusion
trajectories of monomeric phosphoinositide-dependent kinase-1 (PDK1) on a synthetic
target membrane where it can associate with its binding partner protein kinase C alpha isoform (PKC\textalpha) to form a heterodimer detected by its significantly lower diffusivity.

All matlab software is available here: \url{https://github.com/MathBioCU/SingleMolecule}

\section{Introduction}

Inference of physical parameters from experimental data is a ubiquitous
problem in applied mathematics. These parameters can have associated distributions and while there are methods to infer distributions \cite{BanksBortz2005JInverseIll-PosedProbl,MirzaevByrneBortz2016InverseProbl} for the mean-field equation model, it is frequently better to develop a stochastic process model. Building on a stochastic model, a typical aim is to infer the underlying parameters from a realization of the process. For Brownian motion, the parameter of interest is most often the diffusion coefficient. The identification of multiple
diffusion coefficients is the focus of this work in the context of
single particle tracking (SPT). Single particle tracking of proteins
diffusing in 2 dimensions on a biological membrane is a rapidly growing
experimental approach in biophysics and biochemistry. Generally speaking,
it is performed by selectively attaching a small, non-perturbing fluorescent
label (or tag) to a protein of interest, and then recording the fluorescent
protein's movement on the membrane surface using a single molecule
total internal reflection fluorescence (sm TIRF) microscope. The resulting
movie is analyzed to generate a single molecule trajectory suitable
for quantitative analysis \cite{knight2009single}.

In the simplest case, the diffusion coefficient of the process is
inferred via assuming two-dimensional Brownian motion and plotting the mean square displacements of one or more identical protein molecules against time, then calculating the diffusion constant (D) from the slope of the resulting straight line (D = slope / 4). The resulting D value is inversely proportional to the frictional drag of the molecule against the membrane, and thus provides information about the membrane contacts and oligomer number of the molecule as it executes simple Brownian diffusion on a homogeneous membrane surface \cite{knight2009single,falke2010,ziemba2013lateral,ZiembaKnightFalke2012Biochemistry}. However, in more complex systems, Brownian motion can be inadequate for modeling
a protein with multiple diffusive states, since the protein may switch between different membrane docking geometries, or may undergo transient binding interactions with other membrane proteins, cytoskeletal elements, or membrane inhomogeneities \cite{falke2010, ZiembaKnightFalke2012Biochemistry, ziemba2013lateral, BucklesZiembaMassonEtAl2017BiophysicalJournal, BucklesOhashiTremelEtAl2020BiophysicalJournal, Gordon2021, kusumi1993confined, slator2018, schutz1997single, yogurtcu2018cytosolic, ziemba2013lateral, ziemba_falke_2014, saxton1997single, slator2015, qian1991single}.  The resulting deviations of diffusive behavior from simple Brownian motion invite
a deeper interpretation of the underlying dynamics of the system.

The present work focuses on systems in which the protein being observed switches reversibly between two or more protein-membrane or protein-protein interaction states while diffusing on the membrane surface.  When the observed protein forms a new membrane contact, or binds to another membrane-associated protein, 
its lateral diffusion exhibits
a decreased diffusion coefficient owing to the additional frictional
drag of the new membrane contact or bound protein against the lipid bilayer. Quantitative
studies have shown that on supported lipid bilayers, the frictional
drags of multiple species in a complex are additive, enabling calculation
of the diffusion coefficient of the complex from the summed frictional
drags measured for the individual components \cite{falke2010,ziemba2013lateral,ziemba_falke_2014}.
Here we refer to a stochastic diffusion process that includes random switching between a finite number of modes or states, each characterized by distinct
diffusion coefficients, as a ``switch
diffusion process''.

In many biological systems, a protein diffusing on a biological membrane
exhibits multiple interaction states and modes of diffusion with distinct diffusion coefficients.
The protein kinase phosphoinositide-dependent kinase-1 (PDK1) 
possesses a lipid binding domain\footnote{The pleckstrin homology (PH) domain.}
with a strong affinity for specific target lipids\footnote{Phosphatidylinositol (3,4)-bisphosphate ( (PI(3,4)P\textsubscript{2})
and phosphatidylinositol (3,4,5)-trisphosphate PIP\textsubscript{3}} whose levels in the cell membrane are highest during a cell-signaling
event. Single-molecule diffusion studies of full length PDK1 reveal that, while bound to a specific target lipid, PDK1 exhibits
homogeneous 2-D diffusion adequately described by a single diffusion coefficient.
PDK1 also contains a catalytic protein kinase domain which regulates
activation of membrane-bound protein kinase C (PKC\textalpha) \cite{dutil1998regulation,leonard2011crystal}
and other AGC protein kinases. Protein kinase C alpha isoform (PKC\textalpha) is one such kinase that docks to the membrane surface and phosphoactivates
multiple substrate proteins. A single-molecule diffusion study\cite{ziemba_falke_2014}
has revealed that membrane-bound PKC\textalpha\ exhibits two diffusion states distinguished by shallow membrane and deep membrane penetration, as well as correspondingly larger and smaller diffusion coefficients, respectively.
Moreover, PDK1 and PKC\textalpha\ form a stable heterodimeric
complex on the target membrane surface via a PIF interaction between
their kinase domains \cite{Gordon2021}. Given the much greater frictional drag of 
PKC\textalpha\ against the bilayer, single-molecule diffusion analysis has revealed that the binding of PDK1 to PKC\textalpha\  yields a heterodimer in which PDK1 diffusivity is greatly reduced owing to the summed frictional drags of the PKC\textalpha\,and PDK1 molecules \cite{Gordon2021}.

\begin{figure}
\subfloat[The colorbar indicates the frame number.]{\begin{centering}
\includegraphics[width=0.45\columnwidth]{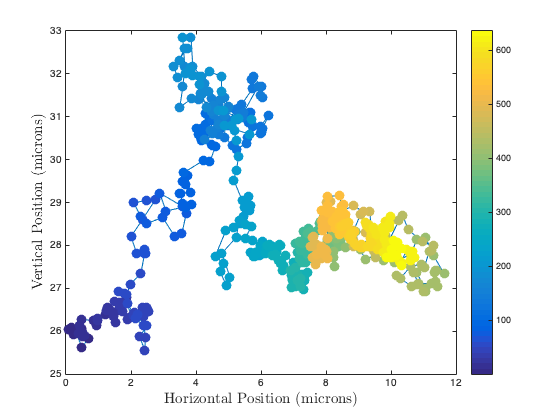}
\par\end{centering}\label{fig:sample_mo7_1a}
}\hfill{}\subfloat[Time series of total displacement between each frame, along with a
30-frame running average (blue curve).]{\begin{centering}
\includegraphics[width=0.45\columnwidth]{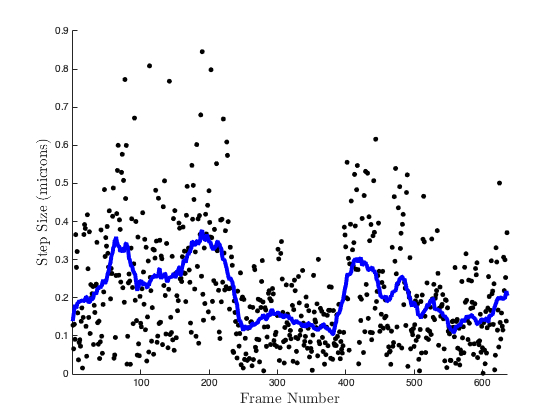}
\par\end{centering}\label{fig:sample_mo7_1b}
}

\caption{A 637-frame trajectory of PDK1 on a synthetic membrane with PKC\textalpha\,concentration
near the $K_{1/2}$ for the binding reaction, ensuring that th average PDK1 molecule is bound to PKC\textalpha\, about half of the time. The observed PDK1 molecule appears to exhibit multiple modes
of diffusion–in particular, the particle displacements
appear to be smaller between frames 230 and 400. The lipid composition was PC/PS/PMA/PIP3 72/24/2/2 mol \%.}
\label{fig:sample_mo7}
\end{figure}

Figure \ref{fig:sample_mo7_1a} depicts the spatial trajectory of a membrane
bound molecule of phosphoinositide-dependent kinase-1 (PDK1) on a synthetic
membrane containing its binding partner protein kinase C (PKC\textalpha). The PKC\textalpha\ concentration is chosen to be near the $K_{1/2}$\footnote{The $K_{1/2}$ is the equilibrium constant obtained in experimental conditions where, on average, half of the protein binding sites are saturated with ligand.}
of the  PDK1-PKC\textalpha\ binding reaction, thereby driving about half of the PDK1 population into PDK1-PKC\textalpha\ heterodimers at any given time.  Figure \ref{fig:sample_mo7_1b} displays step-lengths vs time for consecutive 20 ms steps of the trajectory, as well as a moving average smoothed by averaging the stepsizes over a moving 30-frame window. The latter plot suggests the existence of two or three diffusion states  including a low diffusivity state from frames 230-400, an intermediate state, and a high diffusivity state observed around frames 90, 190 and 410.  

The critical elements of a switch diffusion process are its multiple underlying diffusive states, which may be difficult to directly resolve in the diffusion tracks,
and the random transitions between these states. Two common approaches
to modeling these features are mixture models \cite{Gordon2021,yogurtcu2018cytosolic, falke2010,koo2015extracting}
and hidden Markov models (HMMs) \cite{coombs2009,slator2015,slator2018,linden2018variational,elf2019single,falcao2020diffusion}.
Mixture models are appealing in that they require inference of fewer
parameters, possess parametric statistical forms, and benefit from
the fact that observations may be treated as independent and identically
distributed, which permits use of a joint PDF as a likelihood function \cite{geissen2019inference,bullerjahn2021maximum}.  These models can accurately determine the fraction of total steps associated with each diffusive states, and the diffusion coefficient of each state, but cannot provide kinetic information about the diffusion state lifetimes nor the transition kinetics between states.   HMMs,
by contrast, utilize more parameters and can provide the missing kinetic information. However, observations
from a stochastic process modeled as an HMM are temporally correlated, which invalidates the use of a joint PDF as a likelihood function for parameter inference, which makes parameter inference less analytically tractable.

The advantage of using HMMs to infer physically relevant kinetic information
about the system in question makes research into their implementation
very desirable. Das and colleagues \cite{coombs2009} motivated many
subsequent studies of SPT trajectories using an HMM framework. In
their work, they develop a two-state HMM and describe a Markov Chain
Monte Carlo (MCMC) scheme to infer the parameters of their model and
apply their methodology to study the dynamics of the receptor LFA-1
on the surface of T cells. Moreover, they present
a detailed guide to implementation, including well-annotated
pseudocode of relevant algorithms. We build upon their contribution
by extending their methodology to include a third diffusive state,
opening avenues into methods for statistical model selection and robustness
studies. A contributing author of the Das study, Cairo, extended the
2-state study of LFA-1 to incorporate measurement noise \cite{slator2015},
and implemented a 2-state HMM to detect particle confinement of ganglioside
GM1 lipids to a harmonic potential well \cite{slator2018}. Noise
propagation and confinement are relevant considerations for applied
problems; particle locations can only be measured with finite precision,
and biological membranes exhibit various diffusive characteristics ranging from homogeneous to inhomogeneous. The work presented here focuses on a simple, homogeneous, liquid-disordered membrane lipid bilayer, which is a useful model system for studying signaling reactions on key cellular membranes depleted of lipid microdomains (rafts), including the leukocyte leading edge membrane and the leading edge-derived phagosomal membrane \cite{sitrin2010migrating}.  This simplicity is useful to our initial exploration of the baseline robustness of the extended Das algorithms.  In the present studies, the particle locations are assumed to be known with arbitrary precision, and the membranes employed are flat, well-mixed, and free
of spatially-dependent membrane inhomogeneities and diffusion barriers.  

Lind{é}n and Elf proposed a variational Bayes approach to infer
the underlying parameters of HMMs \cite{persson2013extracting,linden2018variational}.
A significant contribution of this work is a mean-field approximation
to increase the tractability of analysis of their HMM, and a cross-validation
approach to the problem of model selection (that is, identification
of the number of underlying states). Recently, Falcao and Coombs \cite{falcao2020diffusion}
presented an infinite hidden Markov model (iHMM) to simultaneously
predict the cardinality of the state space as well as model parameters.
They implement their method to investigate the motion of receptors
on the plasma membrane of B cells. Absent from both of these studies
is a thorough examination of the robustness of their parameter inference
schemes; these works each validate their methodology with only a handful
of simulated trials.

Predicting from a trajectory (or collection of trajectories) the number
of underlying diffusive states, diffusion coefficients associated
with each state, and characterizing the rate of transition between
different states is a continuing area of research \cite{coombs2009,pohle2017selecting,persson2013extracting,linden2018variational,elf2019single,falcao2020diffusion}.
Robust identification of the number of underlying diffusive states
is a notoriously difficult problem that may provide insight into the
nature of interactions between an objective particle and specific components of its environment.
Likewise, the ability to estimate kinetic parameters of reactions
that give rise to switch diffusion processes open up potential avenues
into researching treatments to related pathologies. For instance,
SPT experiments could be used to screen drug candidates to treat aberrant
signaling behavior, or to study the kinetic impact of mutations on
the signaling or regulatory competence of membrane proteins. Inference
of these parameters and quantification of the robustness of these
inferences are the primary objectives of this work.

In Section \ref{sec:Models-and-Parameter} we will present an inference-motivated
analysis of two common model classes that describe the above phenomena,
as well as a method for inferring Maximum Likelihood Estimates (MLEs)
of the two model classes using Markov chain Monte Carlo (MCMC). In
Section \ref{sec:Simulation-Results} we will empirically compare
the two models using an ensemble of synthetically-produced trajectories
and test the ability of model selection criteria to determine the
number of diffusive states minimal \textit{a priori} information.
Finally, we apply our analysis to a collection of PDK1 trajectories
obtained via single molecule TIRF microscopy, including the trajectory presented
in Figure \ref{fig:sample_mo7}. As part of this analysis, we explore a methodology whereby we treat the diffusion coefficients as known quantities (inferred from previous experimental studies on the individual diffusion states under conditions where switching is not detected \cite{Gordon2021}) in order accelerate and improve estimation of the transition probabilities of the HMM.

Lastly, we note that all matlab software which produces the results is available here: \url{https://github.com/MathBioCU/SingleMolecule}

\section{\label{sec:Models-and-Parameter}Models and Parameter Inference}

Two approaches to modeling switch-diffusion processes are mixture
models and hidden Markov models (HMMs). For the convenience of the
reader and for reference, we include a description of simple Brownian
motion in two spatial dimensions in Section \ref{subsec:One-Diffusive-State}.
The two model classes presented here are differentiated by the random
process that generates the underlying state sequence. The mixture
model state sequence is a Bernoulli scheme, while the HMM state sequence
is a Markov chain. Each provides an accompanying kinetic interpretation.
A Bernoulli scheme describes a zero-order process, since the rate
of product formation is independent of the reactant concentration.
The Markov process, by contrast, describes a first-order reaction,
since the rate of product formation depends on the reactant concentration.

\subsection{Mixture Models\label{subsec:Mixture-Models}}

The simplest model that captures multi-diffusion coefficient motion
is the mixture model, constructed as a sum of Rayleigh distributions, each representing the step size distribution of a single diffusive state with a given diffusion constant D \cite{falke2010}. A particle subject to 2-D Brownian diffusion
whose location is sampled $N$ times at fixed time intervals $\tau$
will produce a sequence of independent 2D displacements $\mathbf{O}=(\mathbf{r}_{i})_{i=1}^{N}$
with with magnitudes $r_{i}=\Vert\mathbf{r}\Vert_{2}$. A particle
subject to $k$ distinct diffusive states $\lbrace D_{i}\rbrace_{i=1}^{k}$
with respective probabilities $\{\alpha_{i}\}_{i=1}^{k}$ has displacements
drawn from the mixture distribution with pdf
\[
f(r;\lbrace\alpha\rbrace_{j}^{k},\lbrace D\rbrace_{j}^{k})=\sum_{j=1}^{k}\alpha_{j}\frac{r}{4D_{j}\tau}e^{-r^{2}/(4D_{j}\tau)},
\]
where the $\alpha_{j}>0$ and $\sum_{j=1}^{k}\alpha_{j}=1$.

MLEs of the $2k-1$ parameters arising from this model $\theta=\lbrace(D_{i})_{i=1}^{k},(\alpha_{i})_{i=1}^{k-1}\rbrace$
can be inferred by maximizing the likelihood function 
\[
L(\theta\vert\mathbf{O})=\prod_{i=1}^{N}f(r_{i}\vert\theta)=\prod_{i=1}^{N}\left(\sum_{i=1}^{k}\alpha_{j}\frac{r_{i}}{4D_{j}\tau}e^{-r^{2}/(4D_{j}\tau)}\right),
\]
or equivalently the log-likelihood function 
\begin{equation}
\ell(\theta\vert\mathbf{O})=\sum_{i=1}^{N}\log\left(\sum_{j=1}^{k}\alpha_{j}\frac{r_{i}}{4D_{j}\tau}e^{-r_{i}^{2}/(4D_{j}\tau)}\right).\label{eqn:mr_log_l-1}
\end{equation}
In contrast to the 1-state model described in the appendix, the MLEs
of this mixture model cannot be computed analytically. In this work,
we calculate the MLE as the maximum a posteriori estimate, obtained
by sampling the posterior distribution of $\theta$ with MCMC. While
MCMC is less computationally efficient than other gradient-based optimization
schemes, it has distinct benefits. Because it simulates a sample of
the posterior distribution of the parameters, it is less liable to
be trapped in local extrema, returning spurious results. Additionally,
simulating the posterior distributions allows quantification of credibility
of parameter estimates.

\subsection{Hidden Markov Models}

\begin{figure}
\centering{}\subfloat{\includegraphics{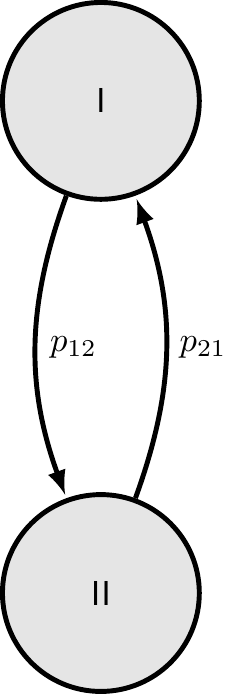}\includegraphics{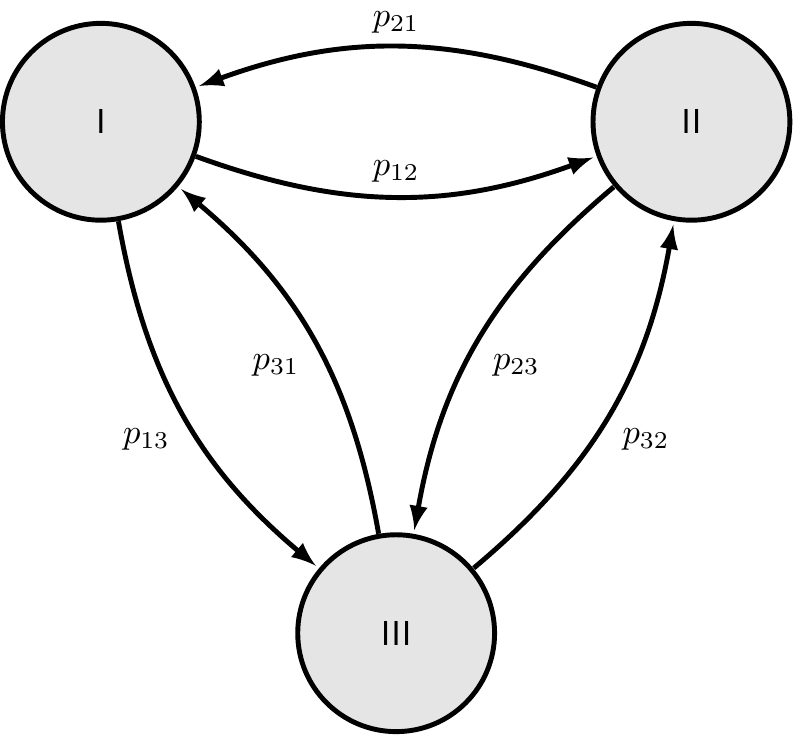}}\caption{A Markov chain permits a description of state switches where the probability
of state transitions depend on the initial state.}
\label{fig:markov}
\end{figure}

The mixture model framework is flexible and general, and is ideal for determining the diffusion coefficients and relative proportions of multiple diffusion states in a complex diffusive system, especially a system with a large number of individual trajectories.  Its limitations include (i) the large number of individual steps that must be measured to define multiple Rayleigh distributions, generally making this model unsuitable for analyzing a individual trajectories, (ii) the inability to determine whether the multiple states detected arise from state-switching within individual trajectories or a mixture of single-state trajectories with distinct diffusion coefficients, and (iii) the lack of information provided about kinetics in a system that switches between states.  

For analyzing individual trajectories in which diffusion switches between distinct states, the HMM can be applied whether or not the likelihood of switching from one state to another depends on the initial state.  This HMM model can reveal the switching kinetics as well as the diffusion coefficients and relative proportions of the states. The observable variables of the HMM are the displacements $\mathbf{O}$ of the particle, and whose hidden
variables $\mathbf{s}=(s_{i})_{i=1}^{N}$, are the states (e.g. ``bound''
or ``unbound'') of the particle during each frame of the trajectory.
See Das et al \cite{coombs2009} for a detailed development of a two-state
hidden Markov model for diffusion. We present the $k$-state generalization
here, though we only consider models with up to $k=3$ diffusive states
in in this work.

In this description, there is a hidden state sequence $\mathbf{s}=(s_{i})_{i=1}^{N}$
with $s_{i}\in\lbrace1,2,...,k\rbrace$. The state sequence $\mathbf{s}$
is a Markov chain with transition matrix $T\in\mathbb{R}^{k\times k}.$
\[
T=\left[\begin{array}{cccc}
p_{11} & p_{12} & ... & p_{1k}\\
p_{21} & p_{22} & ... & p_{2k}\\
\vdots & \vdots & \ddots & \vdots\\
p_{k1} & p_{k2} & ... & p_{kk}
\end{array}\right]
\]
The entries of this transition matrix, $p_{ij}$, are the probabilities
that the particle switches from state $i$ to state $j$ over the
course of a single frame (see Figure \ref{fig:markov} for 2- and
3-state illustrations of the Markov process). While in state $s_{i}$,
the diffusion of the particle is described by diffusion coefficient
$D_{i}$.

This model consists of $k^{2}$ parameters: $k$ diffusion coefficients,
and $k^{2}-k$ independent transition probabilities (making use of
the fact that $T$ is row stochastic, i.e., $\sum_{j}p_{\ell j}=1$).

Because the displacements $r_{i}$ are explicitly not independent,
the likelihood function of a set of parameters $\theta=\lbrace(D_{j})_{j=1}^{k},(p_{j\ell})_{i,\ell=1}^{k}\rbrace$
given an observation $\mathbf{O}$ cannot be succinctly represented
as a joint pdf, as in the mixture model. However, a likelihood function
can be calculated recursively using the algorithm of Baum and Petrie
\cite{bp1969} which we review now.

Begin by considering the quantity 
\[
\alpha_{m}(\kappa)\propto P[r_{1},r_{2},...,r_{m};s_{m}=\kappa|\theta],
\]
which is the probability of observing a partial trajectory $(r_{i})_{i=1}^{m}$
of length $m$ and ending in state $\kappa$, given $\theta$. The
likelihood of observing a full trajectory $\mathbf{O}$ of length
$N$ is the probability of the trajectory ending in any of the $k$
possible states, thus
\begin{equation}
L\left(\mathbf{O}|\theta\right)\propto\sum_{\kappa=1}^{k}{\alpha_{N}(\kappa)}.\label{eqn:hmm_likelihood-1}
\end{equation}
A trajectory of length $m$ that ends in state $\kappa$ can be obtained
from a trajectory of length $m-1$ that transitions into state $\kappa$
from any of the possible states, and then observing a step drawn from
the distribution governing state $\kappa$: 
\begin{align*}
\alpha_{m}(\kappa) & =\underbrace{\left[\sum_{q=1}^{k}\alpha_{m-1}(q)p_{q\kappa}\right]}_{P(\text{transitioning into state }\kappa)}\cdot\underbrace{f(r_{m}\vert s_{m}=\kappa)}_{P(\text{observing displacement }r_{i}\text{ while in state }\kappa)}\\
 & =\left[\sum_{q=1}^{k}\alpha_{m-1}(q)p_{q\kappa}\right]\frac{r_{m}}{4D_{\kappa}\tau}e^{-r_{m}^{2}/(4D_{\kappa}\tau)}.
\end{align*}
This recursion is initialized using the steady-state probabilities
of the Markov chain: $\alpha_{0}(\kappa)=\pi_{\kappa}$. A corresponding
recursive formulation for the log-likelihood is used in practice to
avoid numerical overflow. As with the mixture model of Section \ref{subsec:Mixture-Models},
we calculate the MLE using MCMC.

These two modeling frameworks are closely related, and we note that
the $k$-state mixture model is a special case of the $k$-state hidden
Markov model. A $k$-state mixture model can be recovered by setting
$p_{ij}=\alpha_{j}$ for each $i$, resulting in the transition matrix
\[
T=\left[\begin{array}{cccc}
\alpha_{1} & \alpha_{2} & ... & \alpha_{k}\\
\alpha_{1} & \alpha_{2} & ... & \alpha_{k}\\
\vdots & \vdots & \ddots & \vdots\\
\alpha_{1} & \alpha_{2} & ... & \alpha_{k}
\end{array}\right].
\]
That is, if the transition from state $i$ to $j$ is independent
of state $i$, the resulting HMM is identical to the corresponding
mixture model with the same diffusion coefficients. Conversely, by
the law of large numbers for Markov chains, a sufficiently long trajectory
will produce a distribution of step sizes matching that of a mixture
model with mixing coefficients $\alpha_{i}$ equal to the steady state
distributions $\pi_{i}$ of the Markov chain (provided that the Markov
chain is irreducible).

Kinetic rate constants are related to the transition probabilities through the equation
$$k_{ij} = \frac{p_{ij}}{\tau}\label{eqn:prob_to_rate}$$

\subsection{Parameter Inference via Markov Chain Monte Carlo}

In each of the two models described above, our approach to parameter
inference is leveraged on determining the parameters $\theta_{\text{MLE}}$
that maximize the likelihood (or, equivalently, the log-likelihood)
function of an observation $\mathbf{O}$ given $\theta$. Closed form
expressions for the MLEs (akin to equation \ref{eqn:1draylmle}) do
not exist for the likelihood functions arising from multiple diffusive
states (equations \ref{eqn:mr_log_l-1} and \ref{eqn:hmm_likelihood-1}).
We use Markov chain Monte Carlo (MCMC) to simulate the distribution
of the likelihood function for both classes of models, because the
likelihood functions considered here tend to exhibit multiple local
extrema–especially as the number of underlying states increases–which
can trap gradient-based optimizers. Moreover, because MCMC simulates
the posterior distribution of the likelihood function, the credibility
of the MLEs can be quantified. This in turn allows the robustness
of the parameter estimation scheme for each model class to be directly
compared. We use the maximum a posteriori (MAP) estimate as our MLE.

The three-state HMM presents a difficulty due to the explicit coupling
of the underlying transition probabilities. We aim to infer $(p_{ij})_{i\neq j=1}^{k}$
and deduce $p_{ii}=1-\sum_{j\neq i}p_{ij}$. This requires that $\sum_{j\neq i}p_{ij}\leq1$,
which constrains the domain of the likelihood function to triangular regions in the $p_{ij}p_{ik}$-planes (where $i,j,k$ are distinct). When proposals are generated as perturbations of the previous proposal along the parameter axes, we have observed our MCMC sampler to exhibit unusually high rejection rates, apparently a consequence of the sampler getting ``stuck'' in the corners.

To circumvent this, we use a Matlab implementation of Goodman and
Weare's affine-invariant ensemble MCMC (GWMCMC) \cite{gwmcmc},\cite{emcee}.
Briefly, this implementation uses a parallel ensemble of MCMC chains
(called ``walkers''). New proposals are drawn along the line connecting
two walkers in parameter space. By expanding the space of allowable
directions to draw new samples, GWMCMC mitigates the problem of transition
probabilities getting caught in corners. Moreover, GWMCMC does not
require a researcher to designate the scale of perturbations, since
the magnitude of perturbations scales dynamically as a function of
the distance between walkers in parameter space. In effect, GWMCMC
permits efficient sampling without a priori knowledge of the scales
of the parameters.

In all of the presented results, GWMCMC was run using 200 walkers\footnote{The number of walkers is, aside from the initialization of the ensemble,
the only tunable parameter of GWMCMC. The rule of thumb suggested
by Goodman and Weare is to use a number of walkers that is at least
twice the dimension of the model parameter space, but more tends to
be better \cite{emcee}. We experimented with 50 walker ensembles,
and didn't see a significant speedup or performance increase. On the
other hand, inclusion of more walkers wasn't accompanied by a significant
increase in computational overhead.}. The MCMC chains are initialized randomly, with diffusion coefficients
drawn from a range of $[D_{\text{min}},D_{\text{max}}]$, with 
\[
D_{\text{min}}=\frac{1}{4\tau}\min_{1\leq i\leq N}r_{i}^{2}\text{ and }D_{\text{max}}=\frac{1}{4\tau}\max_{1\leq i\leq N}r_{i}^{2}.
\]
These can be thought of as the largest and smallest point estimates
of a single diffusion coefficient supported by the data. The same
$D_{\text{min}}$ and $D_{\text{max}}$ are used to constrain the
support of the likelihood function; in absence of this constraint,
the MCMC sampler occasionally diverges, as an unreasonably large diffusion
coefficient can be offset in likelihood by a Markov chain that seldom
visits the anomalous state. Transition rates are initialized entirely
randomly, constrained only by the stochasticity property of the transition
matrix (as described above). The remaining 199 walkers are initialized
in a tight Gaussian ball around the first walker.

\subsubsection{Convergence of MCMC Sampler\label{sec:mcmc_convergence}}

Standard MCMC convergence diagnostics (such as the Gelman-Rubin statistic)
cannot be implemented in applications of GWMCMC because the parallel
chains of the Goodman-Weare method are not independent. Instead, the
standard approach to assessing convergence is to estimate the integrated
autocorrelation time, $\tau_{\text{{corr}}}$, of the chains, which
can be used to quantify the variance of the sampler. This can be leveraged
to determine how many steps are necessary to reduce the relative error
of the estimated MLE to be suitably small\cite{emcee},\cite{sokal1996}.
Specifically, the criteria for convergence is

\[
z_{\gamma}\sqrt{{\frac{{\sigma_{\text{{\ensuremath{\theta}}}}^{2}}}{M_{\text{{eff}}}}}}\leq\hat{{\mu}}_{\theta}\beta,
\]

where $z_{\gamma}$ is the $z$-score, or ``critical value'' associated
with a confidence level $\gamma$, $\sigma_{\theta}^{2}$ is the sample
variance associated with samples of the parameter $\theta$, $M_{\text{{eff}}}$
is the effective independent sample size of the chain ($M_{\text{{eff}}}=\frac{{M}}{\tau_{\text{{corr}}}}$),
$\hat{\mu}_{\theta}$ is the sample mean of the parameter $\theta$,
and $\beta$ is the desired relative error of the estimated MLE. The
chains were diagnosed as converged when this inequality held for each
parameter $\theta$. Essentially, the expression on the left is the
width of a $\gamma$-level confidence interval based on effectively
independent samplings of $\theta$. The expression on the right is
a threshold for the width of the confidence interval, relative to
the sample mean of the parameter. In this work, we selected $\gamma=0.95$
(so $z_{\gamma}\approx1.96$) and $\beta=10^{-4}.$



\section{\label{sec:Simulation-Results}Model Comparisons and Robustness Results}

In this section, we use an ensemble of simulated trajectories to empirically
compare the robustness of the inference scheme described above across
model types. As part of this analysis, we investigate how the robustness
of the method scales with the number of underlying states, giving
particular focus to parameter estimation of three-state models. Section
\ref{subsec:Simulation-Design} describes our procedure for simulating
trajectories in the ensemble. Section \ref{subsec:Comparison-of-HMM}
illustrates the inference scheme for a particular trajectory in the
ensemble. Results of the full ensemble analysis is contained in Section
\ref{subsec:Ensemble-Robustness-Results}, which includes an application
of model selection criteria to predict the number of underlying diffusive
states. Finally, we apply our analysis to a collection of experimentally
measured trajectories of PDK1 in Section \ref{subsec:Parameter-Inference-from}.

\subsection{Simulation Design\label{subsec:Simulation-Design}}

The numerical experiment that follows uses synthetic diffusion data
generated in the following way: it is assumed that the diffusive
state transitions are described by a Markov process. To this end,
a random (entirely arbitrary) $3\times3$ transition matrix is constructed.
Next, a set of diffusion coefficients is randomly selected from a
range of ``biologically feasible'' parameters, summarized in Table
\ref{tab:feasible_params}. These were chosen to be roughly compatible
with estimations of diffusion coefficients from previous work considering
membrane-targeting proteins \cite{falke2010}.
\begin{center}
\begin{table}[ht]
\centering{}%
\begin{tabular}{ccc}
$D_{1}$ & $D_{2}$ & $D_{3}$\tabularnewline
\midrule 
$(10^{-3},10^{-2})$ & $(10^{-2},10^{-1})$ & $(0.5,5)$\tabularnewline
\end{tabular}\caption[Biologically feasible diffusion coefficients used to simulate trajectories]{Ranges for ``biologically feasible'' diffusion coefficients. Units
are $\mu$m\protect\protect\textsuperscript{2}/s.}
\label{tab:feasible_params}
\end{table}
\par\end{center}

The trajectory was initialized randomly with the probability of initializing
in state $i$ equal to the stationary probability of state $i$ of
the transition matrix. The particle's displacement was obtained by
drawing coordinate displacements $x$ and $y$ from $\mathcal{N}(0,2D\tau)$.
Subsequent states were determined by propagating the Markov chain,
which dictated subsequent coordinate displacements. This process was
repeated until a trajectory consisting of 1000 displacements was obtained.

\subsection{Comparison of HMM and Mixture Models\label{subsec:Comparison-of-HMM}}

Consider a particle subject to three diffusive states\footnote{We illustrate our analysis for a 3-state model. A 2-state analysis is similar, see \cite{coombs2009}} in two spatial
dimensions, with displacements measured for 1000 frames\footnote{This is a commonly selected movie length in the HMM literature for
simulated single particle trajectories, as seen in \cite{coombs2009},
\cite{slator2015}, \cite{slator2018}.} with frame lag $\tau=0.01$. In order of increasing magnitude, the
diffusion coefficients corresponding to each state were selected to
be $D_{1}=0.0036513\mu\text{m}^{2}/\text{s},D_{2}=0.024517\mu\text{m}\textsuperscript{2}/\text{s},D_{3}=1.6539\mu\text{m}^{2}/\text{s}$.

The underlying state sequence was generated from the transition matrix
\[
T=\left[\begin{array}{ccc}
0.3311 & 0.4693 & 0.1997\\
0.9023 & 0.0716 & 0.0261\\
0.3164 & 0.0554 & 0.6282
\end{array}\right].
\]
The expected values of these mixture coefficients obtained from the
steady state of the Markov chain are $\alpha_{1}=0.473$, $\alpha_{2}=0.2552$,
and $\alpha_{3}=0.2719$.

For the purpose of this analysis, we wished to differentiate bias
in parameter inference due to the inference scheme from the bias arising
from taking a finite sample of a random process (i.e., sampling error).
Thus, we calculated transition rates and mixing coefficients directly
from the state sequence of the simulated trajectory to obtain 
\[
T=\left[\begin{array}{ccc}
0.3453 & 0.4691 & 0.1856\\
0.8989 & 0.0903 & 0.0108\\
0.3529 & 0.0769 & 0.5701
\end{array}\right]
\]
and mixing coefficients $\alpha_{1}=0.501,\alpha_{2}=0.277,\alpha_{3}=0.222$.
Likewise, each diffusion coefficient was estimated from the sample
using Equation \ref{eqn:1draylmle}, using the subset of displacements
corresponding to each state. In other words, we calculated the best
possible estimate of each diffusion coefficient as if we had perfect
knowledge of the particle's state sequence. The parameters used to
generate the sample are tabulated along with these retroactively calculated
parameters in Table \ref{tab:pop_params}. Throughout this work, when
we compare an inferred parameter to a true value, we use the parameters
computed from the sample.
\begin{center}
\begin{table}[htb]
\centering{}{\small{}}%
\begin{tabular}{lccccccccccc}
 & {\small{}$D_{1}$} & {\small{}$D_{2}$} & {\small{}$D_{3}$} & {\small{}$p_{12}$} & {\small{}$p_{13}$} & {\small{}$p_{21}$} & {\small{}$p_{23}$} & {\small{}$p_{31}$} & {\small{}$p_{32}$} & {\small{}$\alpha_{1}$} & {\small{}$\alpha_{2}$}\tabularnewline
\midrule 
{\small{}G} & {\small{}0.0089} & {\small{}0.0702} & {\small{}1.4083} & {\small{}0.4693} & {\small{}0.1997} & {\small{}0.9023} & {\small{}0.0261} & {\small{}0.3164} & {\small{}0.0554} & {\small{}0.473} & {\small{}0.255}\tabularnewline
{\small{}S} & {\small{}0.0088} & {\small{}0.0716} & {\small{}1.5516} & {\small{}0.4691} & {\small{}0.1856} & {\small{}0.8989} & {\small{}0.0108} & {\small{}0.3529} & {\small{}0.0769} & {\small{}0.501} & {\small{}0.277}\tabularnewline
\end{tabular}{\small{}\caption[Ground truth parameters in simulated trajectory]{Parameters used in simulation comparing HMM and mixture model classes.
Diffusion coefficients are reported in units of $\mu\text{m}^{2}/\text{s}$.
The row of generated ``G'' parameters are those used to generate
the trajectory, the row of sample ``S'' parameters are those retroactively
calculated from perfect knowledge of the trajectory.{\small{}\label{tab:pop_params}}}
}
\end{table}
\par\end{center}

For the 3-state HMM, we performed a 200 walker GWMCMC sampling of
the posterior distributions until the convergence criterion described
in Section \ref{sec:mcmc_convergence} was achieved, which occurred
after $1.7\times10^{7}$ total steps, or $85,000$ steps per walker\footnote{The convergence criterion was checked after every $10^{5}$ steps.}.
Figure \ref{fig:3state_sample_traces} depicts a trace of one walker
in each parameter (other walkers of the ensemble are similar). The
(sample) parameter that we aim to infer is indicated by a horizontal
black line. Indeed, these traces appear to be reasonably well converged
to the target value.
\begin{center}
\begin{figure}[!htb]
\centering{}\includegraphics[width=0.4\textwidth]{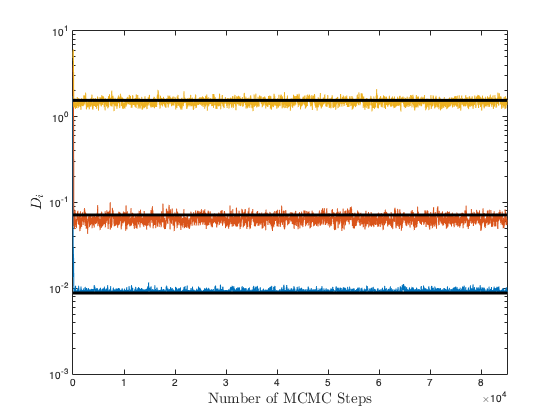}\\
 \includegraphics[width=0.3\textwidth]{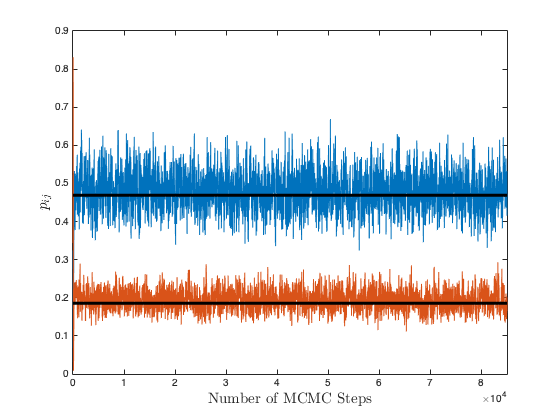}
\includegraphics[width=0.3\textwidth]{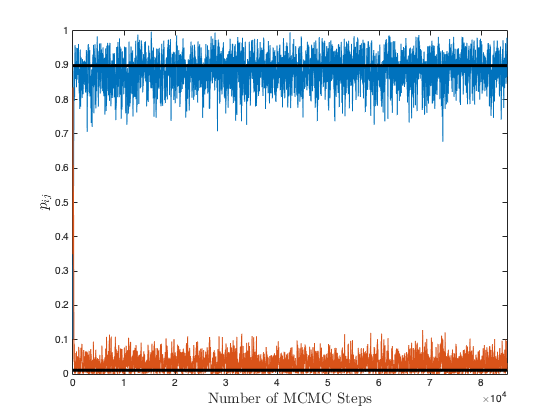}
\includegraphics[width=0.3\textwidth]{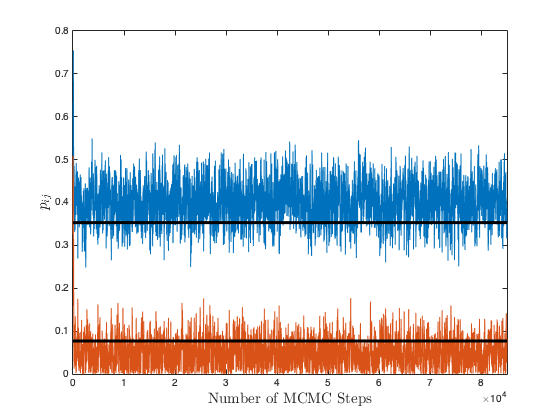}
\caption[3-state HMM trace]{3-state HMM parameter estimation traces. Top: Diffusion coefficients,
Bottom left; $p_{12}$ and $p_{13}$, Bottom center: $p_{21}$ and
$p_{23}$, Bottom right: $p_{31}$ and $p_{32}$.}
\label{fig:3state_sample_traces}
\end{figure}
\par\end{center}

\begin{figure}[!htbp]
\centering{}\subfloat[Diffusion coefficients.]{\includegraphics[height=0.2\textheight]{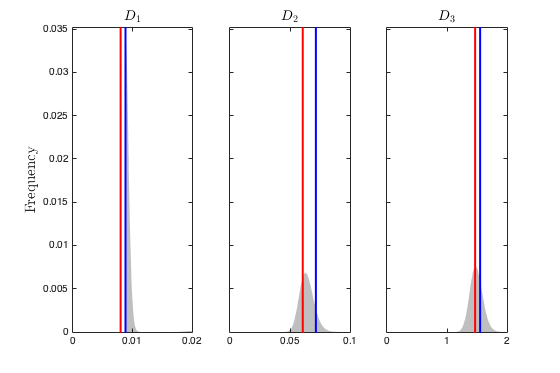}\quad{}

}\subfloat[Transition probabilities.]{\includegraphics[height=0.2\textheight]{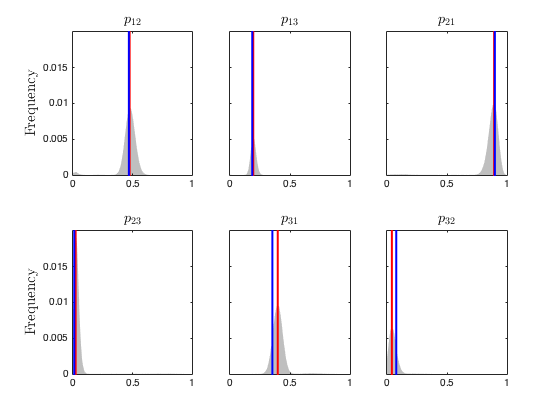}

}\caption[3-state HMM posterior distributions from sample ensemble trajectory]{Posterior distributions of parameters obtained from a 3-state HMM
analysis.}
\label{fig:hmm3_posterior_sampletraj}
\end{figure}

The marginalized posterior distributions\footnote{Because we use a uniform prior, the posterior distributions sampled
by GWMCMC are proportional to the likelihood function.} obtained from our sampler are depicted in Figure \ref{fig:hmm3_posterior_sampletraj}.
Vertical axis labels are omitted because the distributions are proportional
to the likelihood function, but the particular heights of the distributions
are irrelevant to our analysis. True (sample) parameter values are
indicated by vertical blue bars. The mode of the distribution is taken
to be our MLE\footnote{Maximum \textit{a posteriori} (MAP) estimate is also used to describe
this choice of estimator. The two are equivalent when using a uniform
prior.}, which is marked by a vertical blue line. The mode of the distribution
is identified from a histogram of the GWMCMC samples, indicated in
red. The value of the mode depends on how the data is binned into
a histogram\footnote{In order to clearly depict the posterior distributions, they are depicted
on an interval that excludes rare, unusually large samples. When the
mode of the distribution is calculated, these observations are included,
which changes the width of the histogram bins. This is why the calculated
mode (red line) doesn't always align with the apparent mode of the
histogram.}; the default number of bins selected to discretize posterior distributions
in this and subsequent analyses was 1000. We used the same binning
to calculate the highest density posterior interval (HDPI) containing
95\% of the posterior density, which we report in Table \ref{tab:sample_trial_parameter_estimates}
along with the MLE for each parameter. This credible interval quantifies
the uncertainty of our MLE.

A similar analysis was performed for a 3-state mixture model using
the same simulated trajectory. We omit the traces and posterior distributions
for brevity. The GWMCMC sampler achieved convergence after $8.6\times10^{5}$
total steps, or 43000 steps per walker. MLEs of diffusion coefficients
and mixing coefficients obtained from this analysis are tabulated
in Table \ref{tab:sample_trial_parameter_estimates}, along with their
corresponding 95\% credible intervals.

\begin{figure}[!htbp]
\centering{}\centering \includegraphics[width=0.4\textwidth]{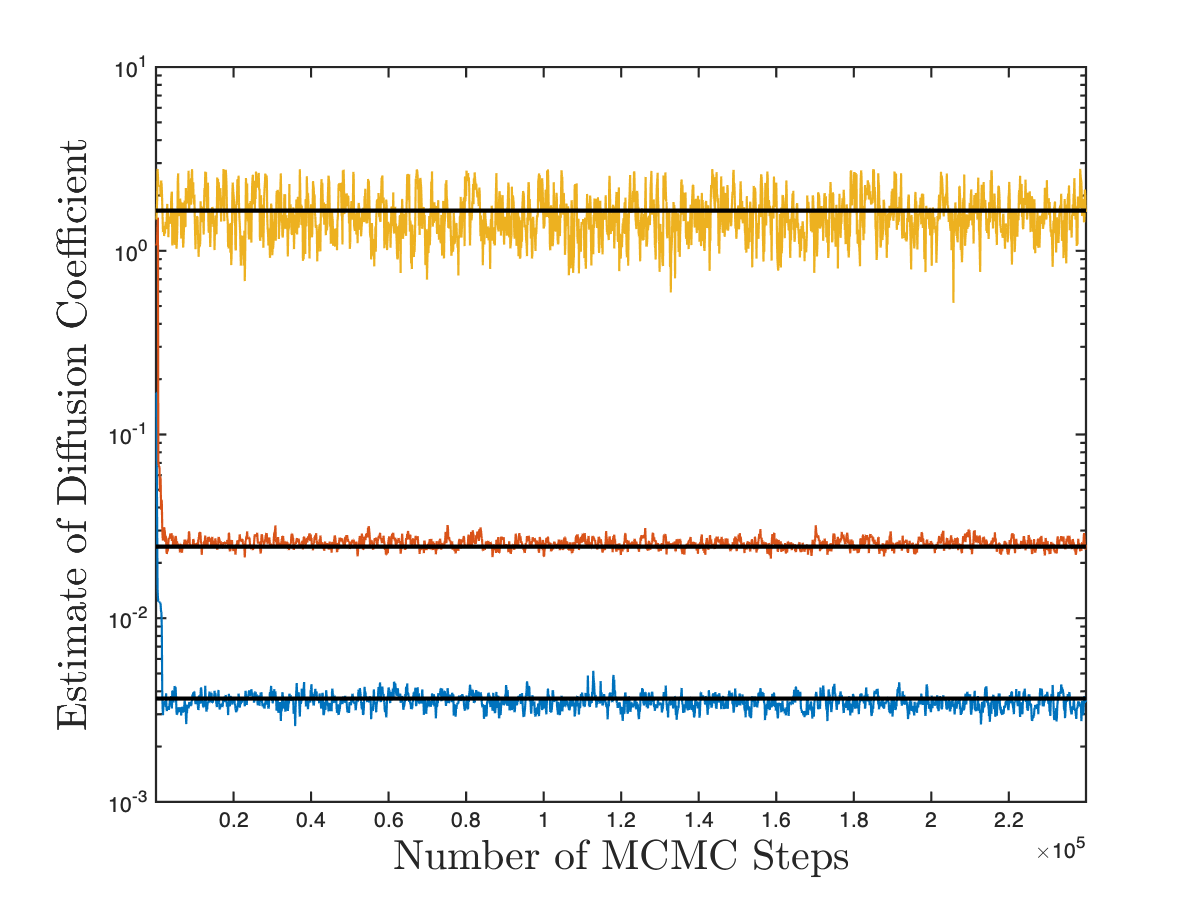}
\includegraphics[width=0.4\textwidth]{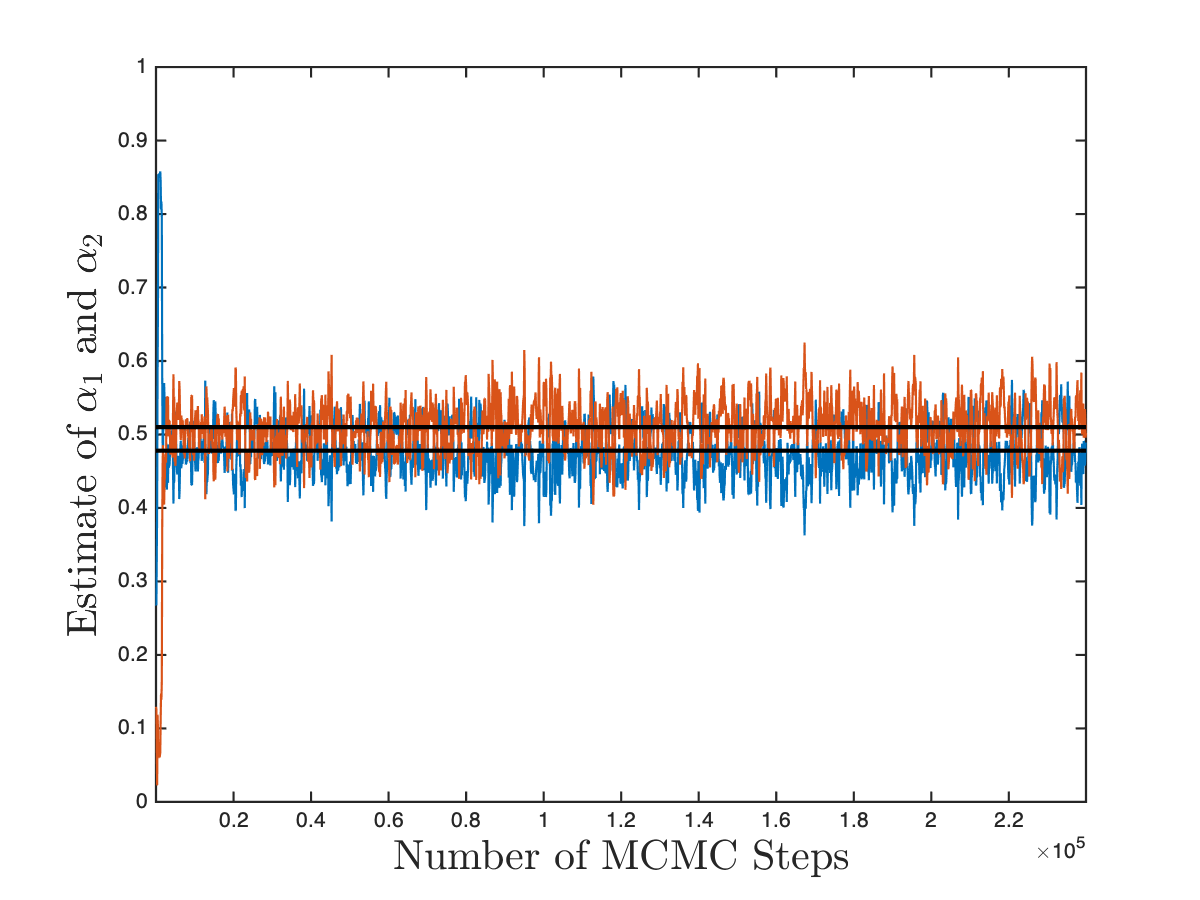}
\caption[3-state mixture model trace]{3-state mixture model parameter estimation traces. True parameter
values are indicated with a thick black line. Left: MCMC samples of
diffusion coefficients; Right: MCMC samples of mixture coefficients.}
\label{fig:3state_sample_traces_mr}
\end{figure}

\begin{table}[!htb]
\centering{}\centering \caption{Summary of parameter estimation results for the trial described above.}
\begin{tabular}{cccrccrc}
 & $\theta$ & $\hat{\theta}_{\text{mix}}$ & 95\% CI & $\frac{\vert\theta-\hat{\theta}\vert}{\theta}$ & $\hat{\theta}_{\text{HMM}}$ & 95\% CI & $\frac{\vert\theta-\hat{\theta}_{\text{HMM}}\vert}{\theta}$\tabularnewline
\midrule 
$D_{1}$ & 0.0088 & 0.0084 & {[}0.0066,0.0097{]} & 0.0502 & 0.0080 & {[}0.0064,0.0191{]} & 0.0926\tabularnewline
$D_{2}$ & 0.0716 & 0.0565 & {[}0.006,0.0725{]} & 0.2104 & 0.0607 & {[}0.0424,1.1757{]} & 0.1530\tabularnewline
$D_{3}$ & 1.5516 & 1.4719 & {[}1.2784,1.7337{]} & 0.0514 & 1.4694 & {[}1.2644,1.7153{]} & 0.0530\tabularnewline
$\alpha_{1}$ & 0.501 & 0.4738 & {[}0.4052,0.5396{]} & 0.0542 & 0.4916 & [0.4473, 0.5382] & 0.0187 \tabularnewline
$\alpha_{2}$ & 0.277 & 0.2935 & {[}0.2338,0.3546{]} & 0.0594 & 0.2709 & [0.2271, 0.3191] & 0.0220\tabularnewline
$p_{12}$ & 0.4691 & - & - & - & 0.4762 & {[}0.0155,0.5642{]} & 0.0151\tabularnewline
$p_{13}$ & 0.1856 & - & - & - & 0.1935 & {[}0.1519,0.2415{]} & 0.0424\tabularnewline
$p_{21}$ & 0.8989 & - & - & - & 0.8930 & {[}0.3100,0.9850{]} & 0.0066\tabularnewline
$p_{23}$ & 0.0108 & - & - & - & 0.0220 & {[}0,0.5384{]} & 1.0291\tabularnewline
$p_{31}$ & 0.3529 & - & - & - & 0.3976 & {[}0.2977,0.6014{]} & 0.1266\tabularnewline
$p_{32}$ & 0.0769 & - & - & - & 0.0409 & {[}0,0.1790{]} & 0.4679\tabularnewline
\end{tabular}\label{tab:sample_trial_parameter_estimates}
\end{table}

Broadly, diffusion coefficients are estimated with a fairly high degree
of accuracy and neither model performs uniformly better than the other.
$D_{2}$ is both the least accurate and least certain estimate for
both models. A plausible explanation is that the $D_{2}$ serves as
a degree of freedom within the model to describe unusually large displacements
arising from state 1, or unusually small displacements arising from
state 3. Mixture coefficients are inferred with high accuracy and
moderate precision. Transition probabilities of the HMM, by contrast,
are more biased than than the mixing coefficients. Curiously, the
MLE transition rates give rise to a steady state distribution that
is more accurate than that inferred through the mixture model: $\alpha_{1}=0.4976,\alpha_{2}=0.2694$.

In conclusion, neither model seems to decisively outperform the other
for this trajectory. The HMM took significantly longer to run to convergence
(roughly 18-fold longer), and this extended computational time doesn't
directly translate to more accurate parameter inference. However,
the advantage of an HMM as a kinetic model may, for some applications,
outweigh the computational cost. In an effort to explore how well
these results hold up in general, we tested this procedure on the
full ensemble of simulated trajectories.

\subsection{Ensemble Robustness Results Using Simulated Trajectories\label{subsec:Ensemble-Robustness-Results-1}}

\label{subsec:Ensemble-Robustness-Results}

\begin{figure}[!htb]
\begin{centering}
\includegraphics[width=0.6\textwidth]{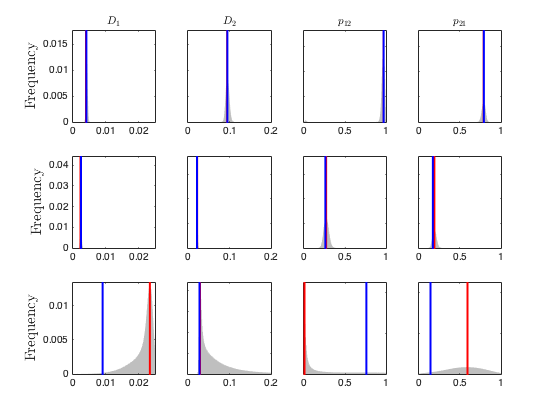}
\par\end{centering}
\caption[Representative 2-state HMM posteriors from ensemble robustness analysis]{Posterior distributions of 2-state HMM parameters from synthetic
data. The first row depicts a trial where parameters are inferred
with the highest accuracy (minimum $L^{1}$ relative error); the second
row depicts a typical trial (median $L^{1}$ relative error); the
third row depicts a trial where parameters are inferred with worst
accuracy (maximum $L^{1}$ relative error).}
\label{fig:sample_2state_posteriors}
\end{figure}

\begin{figure}[!htbp]
\centering{}%
\begin{minipage}[t]{0.4\columnwidth}%
\begin{center}
\includegraphics[scale=0.4]{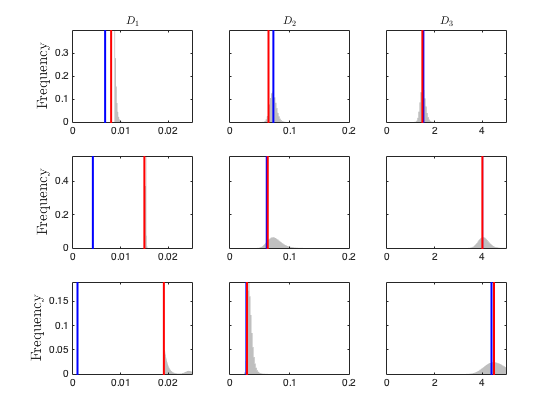}
\par\end{center}%
\end{minipage}\hfill{}%
\begin{minipage}[t]{0.4\columnwidth}%
\begin{center}
\includegraphics[scale=0.4]{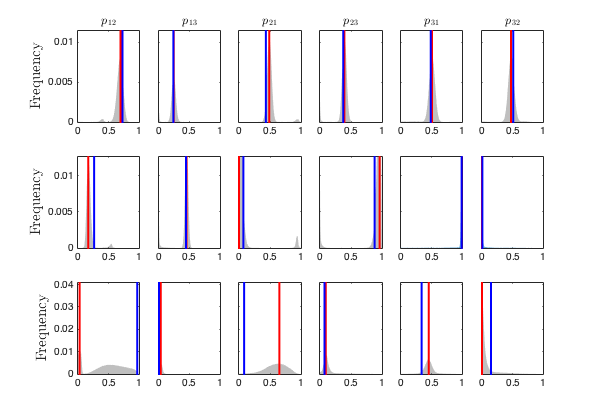}
\par\end{center}%
\end{minipage}\caption[Representative 3-state HMM posteriors from ensemble robustness analysis]{Posterior distributions of 3-state HMM parameters from synthetic
data. The first row depicts a trial where parameters are inferred
with the highest accuracy (minimum $L^{1}$ relative error); the second
row depicts a typical trial (median $L^{1}$ relative error); the
third row depicts a trial where parameters are inferred with worst
accuracy (maximum $L^{1}$ relative error).}
\label{fig:sample_3state_posteriors}
\end{figure}

\begin{figure}[!htbp]
\begin{centering}
\begin{minipage}[t]{0.45\columnwidth}%
\begin{center}
\includegraphics[scale=0.4]{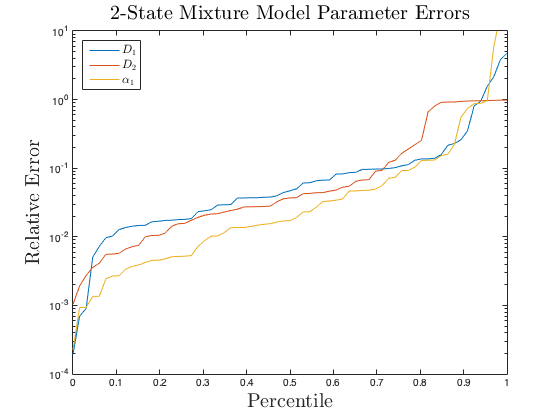}
\par\end{center}%
\end{minipage}\hfill{}%
\begin{minipage}[t]{0.45\columnwidth}%
\begin{center}
\includegraphics[scale=0.4]{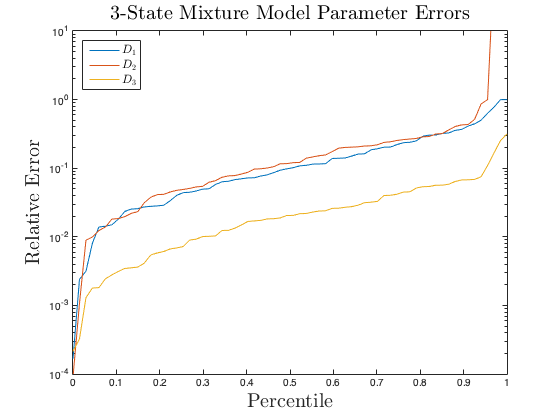}
\par\end{center}%
\end{minipage}
\par\end{centering}
\centering{}%
\begin{minipage}[t]{0.45\columnwidth}%
\begin{center}
\includegraphics[scale=0.4]{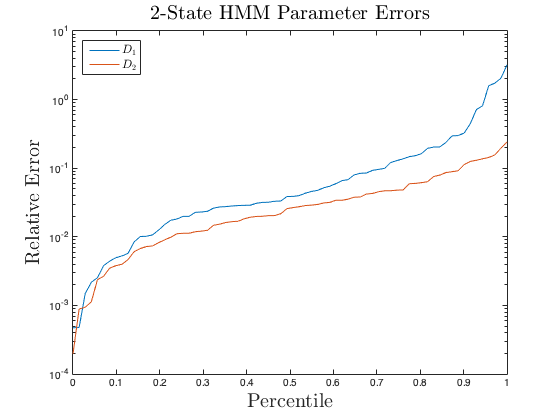}
\par\end{center}%
\end{minipage}\hfill{}%
\begin{minipage}[t]{0.45\columnwidth}%
\begin{center}
\includegraphics[scale=0.4]{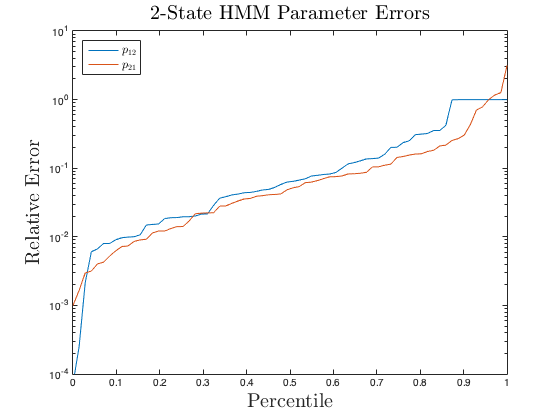}
\par\end{center}%
\end{minipage}
\begin{minipage}[t]{0.45\columnwidth}%
\begin{center}
\includegraphics[scale=0.4]{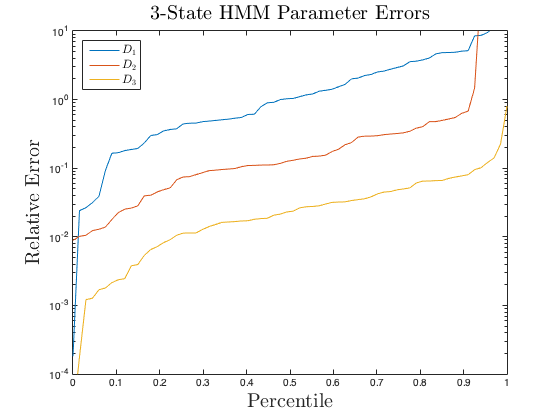}
\par\end{center}%
\end{minipage}\hfill{}%
\begin{minipage}[t]{0.45\columnwidth}%
\begin{center}
\includegraphics[scale=0.4]{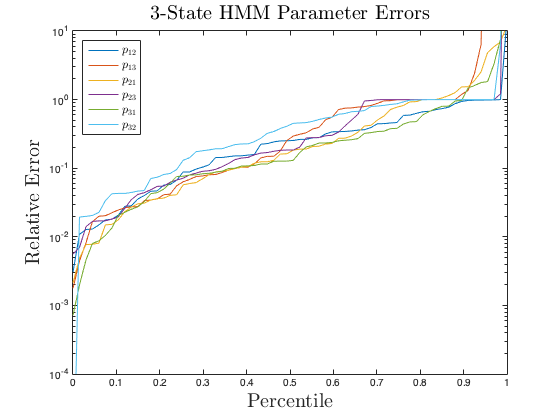}
\par\end{center}%
\end{minipage} \caption[Relative error from simulated trajectories]{Relative error of 2- and 3-state HMM parameter estimates from ensemble of synthetic trajectories.}
\label{fig:relative_error_plots}
\end{figure}

The example presented in the previous section illustrates a situation where the parameter inference scheme works well for both models. In pursuit of a characterization of the relative robustness of the two methods, as well as the relationship between robustness and the dimension of parameter space, we performed a similar analysis on the remainder of the ensemble described in Section \ref{subsec:Simulation-Design}. From each realization, we implemented the inference scheme described in Section \ref{sec:Models-and-Parameter} to obtain 2- and 3-state model parameter estimates for both the mixture model and the HMM. We calculated the diffusion coefficient of a 1-state model via Equation \ref{eqn:1draylmle}. The estimated parameters were compared to the sample parameters inferred from perfect knowledge of the particle's state sequence, as in the previous section. Figures \ref{fig:sample_2state_posteriors} and \ref{fig:sample_3state_posteriors} each illustrate the posterior distributions of HMM parameters for three sample trials representative of poor, median, and good trials (based on the $L^{1}$ relative error of the parameter estimates). The relative errors for all trials are depicted in Figure \ref{fig:relative_error_plots}, which plots the relative error of each parameter for each trial. Tables \ref{tab:mr2_bias_summary}-\ref{tab:hmm3_bias_summary} summarize the relative bias across the ensemble for each model class. For each model, the quartiles of the relative absolute bias describe typical error among the model parameters across the ensemble trajectories. The uncertainty of these estimate is summarized in these tables by the ``width ratio''\footnote{The width ratio is calculated for parameter $\theta$ with MLE $\hat{\theta}$
and 95\% credible interval $[\theta_{-},\theta_{+}]$ as $\max(\lbrace\vert\hat{\theta}-\theta_{-}\vert,\vert\hat{\theta}-\theta_{+}\vert)/\hat{\theta}$.
It is the ratio of the distance from the MLE to the furthest edge
of the credible interval and the MLE itself.}, which illustrates the relative uncertainty in the estimate for each parameter.

Unsurprisingly, the 2-state model parameters are generally inferred with greater fidelity than the 3-state models. Both the 2-state mixture model and HMM parameters are typically inferred within about 6\% of the sample value. The 3-state mixture model parameters are typically inferred within 12\% of the sample value. By contrast, the inference scheme struggles has difficulty inferring the smaller diffusion coefficients of the 3-state HMM ($D_1$ is typically inferred with 100\% error), and 20\% error is typical among the transition probabilities. This disparity is due in part to the fact that fewer than a third of the 3-state HMM analyses met the convergence criterion, highlighting the computational challenge associated with this analysis.

\begin{table}[!htb]
\begin{centering}
\begin{tabular}{cccccccc}
 &  & Relative (Absolute) Bias &  &  &  & Width Ratio & \tabularnewline
 & 25\% & 50\% & 75\% &  & 25\% & 50\% & 75\%\tabularnewline
\hline 
$D_{1}$ & 0.0190 & 0.0388 & 0.1331 &  & 0.2239 & 0.3429 & 0.0764\tabularnewline
$D_{2}$ & 0.0011 & 0.0262 & 0.0480 &  & 0.1343 & 0.2146 & 0.1541\tabularnewline
$\alpha_{1}$ & 0.0193 & 0.0635 & 0.2190 &  & 0.1506 & 0.3067 & 0.5598\tabularnewline
\end{tabular}
\par\end{centering}
\caption[Summary of 2-state mixture model inference robustness]{Quartiles of relative bias and width ratio among the estimates from
2-state mixture trials. These statistics summarize 71 independent
trials, of which 70 met the convergence criterion.}

\label{tab:mr2_bias_summary}
\end{table}

\begin{table}[!htb]
\begin{centering}
\begin{tabular}{cccccccc}
 &  & Relative (Absolute) Bias &  &  &  & Width Ratio & \tabularnewline
 & 25\% & 50\% & 75\% &  & 25\% & 50\% & 75\%\tabularnewline
\hline 
$D_{1}$ & 0.0182 & 0.0331 & 0.0962 &  & 0.1227 & 0.1717 & 0.3071\tabularnewline
$D_{2}$ & 0.0113 & 0.0262 & 0.0482 &  & 0.1174 & 0.1470 & 0.2957\tabularnewline
$p_{12}$ & 0.0190 & 0.0507 & 0.1379 &  & 0.1500 & 0.2498 & 0.7274\tabularnewline
$p_{21}$ & 0.0131 & 0.0449 & 0.1107 &  & 0.1052 & 0.2273 & 0.3940\tabularnewline
\end{tabular}
\par\end{centering}
\caption[Summary of 2-state HMM inference robustness]{Quartiles of relative bias and width ratio among the estimates from
2-state HMM trials. These statistics summarize 66 independent trials.}

\label{tab:hmm2_bias_summary}
\end{table}

\begin{table}[!htb]
\begin{centering}
\begin{tabular}{cccccccc}
 &  & Relative (Absolute) Bias &  &  &  & Width Ratio & \tabularnewline
 & 25\% & 50\% & 75\% &  & 25\% & 50\% & 75\%\tabularnewline
\hline 
$D_{1}$ & 0.0421 & 0.0998 & 0.2281 &  & 0.3422 & 0.4523 & 0.7342\tabularnewline
$D_{2}$ & 0.0483 & 0.1185 & 0.2574 &  & 0.3126 & 0.5085 & 1\tabularnewline
$D_{3}$ & 0.0070 & 0.0205 & 0.0432 &  & 0.1476 & 0.1825 & 0.2460\tabularnewline
$\alpha_{1}$ & 0.0131 & 0.0322 & 0.0706 &  & 0.0965 & 0.1442 & 0.1917\tabularnewline
$\alpha_{2}$ & 0.0162 & 0.0341 & 0.807 &  & 0.1112 & 0.1431 & 0.2490\tabularnewline
\end{tabular}
\par\end{centering}
\centering{}\caption[Summary of 3-state mixture model inference robustness]{Quartiles of relative bias and width ratio among the estimates from
3-state mixture model trials. These statistics summarize 68 independent
trials, 46 of which met convergence criteria after 24 hours (wall-clock
time).\label{tab:mr3_bias_summary}}
\end{table}

\begin{table}[!htb]
\begin{centering}
\begin{tabular}{cccccccc}
 &  & Relative (Absolute) Bias &  &  &  & Width Ratio & \tabularnewline
 & 25\% & 50\% & 75\% &  & 25\% & 50\% & 75\%\tabularnewline
\hline 
$D_{1}$ & 0.3043 & 1.0197 & 2.9995 &  & 0.3049 & 0.4693 & 1.1512\tabularnewline
$D_{2}$ & 0.0514 & 0.1405 & 0.4244 &  & 0.3547 & 0.6628 & 51.40\tabularnewline
$D_{3}$ & 0.0092 & 0.0247 & 0.0502 &  & 0.1862 & 0.2838 & 0.6269\tabularnewline
All $p_{ij}$ & 0.0647 & 0.2234 & 0.8266 &  & 0.2755 & 0.7621 & 1.757\tabularnewline
\end{tabular}
\par\end{centering}
\caption[Summary of 3-state HMM inference robustness]{Quartiles of relative bias and width ratio among the estimates from
3-state HMM trials. Transition rates are pooled into the category
of ``all $p_{ij}$''. These statistics summarize 68 independent
trials, 22 of which met convergence criteria after 24 hours (wall-clock
time).}

\label{tab:hmm3_bias_summary}
\end{table}

Convergence and duration of the GWMCMC sampler is summarized in Figure \ref{fig:mcmc_convergence_plots}. Duration is quantified by the number of samples per walker. Among the 200 trajectories of the ensemble 154 trajectories converged for the 3-state HMM analysis, 177 trajectories converged for the 3-state mixture model, 197 trajectories converged for the 2-state HMM, and 197 trajectories converged for the 2-state mixture model.

\subsubsection{Number of States Determined by Model Selection Criteria}

The Akaike information criterion (AIC) is computed as 
\[
\text{AIC}=2k-2\hat{\ell}
\]
where $k$ is the number of estimated parameters and $\hat{\ell}$
is the maximum value of the log-likelihood function for the model.

The Bayes information criterion (BIC), on the other hand, is derived
from a Bayesian viewpoint with equal prior probability on each model
and vague priors (the parameters are from a distribution from the
Koopman-Darmois family, a very general parametric family) \cite{schwarz1978}
is computed as 
\[
\text{BIC}=\log(n)k-2\hat{\ell}
\]
with $k,\hat{\ell}$ as defined above and sample size $n$. Common
to the AIC and BIC is a term to penalize complex models and a goodness-of-fit
term to select models that more accurately reflect an observed sample.
Specifically, the BIC assigns a more significant penalty to parameter
values when the sample size is larger.

Using the MLEs obtained from the ensemble, we computed the AIC and
BIC for each model type, for each trajectory of the ensemble. The
model with the lowest score in each criteria was selected as the most
suitable model for the data. The AIC and BIC both performed well based
on HMM parameter estimates (81.5\% and 83\% accuracy, respectively).
The AIC and BIC based on mixture model parameter estimates performed
less well by comparison (73.5\% and 72.5\% accuracy, respectively).
\begin{center}
\begin{table}[htb]
\begin{centering}
\begin{tabular}{ccc}
 & AIC & BIC\tabularnewline
\midrule 
HMM & 81.5\% & 83\%\tabularnewline
Mixture & 73.5\% & 72.5\%\tabularnewline
\end{tabular}
\par\end{centering}
\caption[Robustness of State Predictions]{Accuracy of AIC and BIC as a means to predict the number of underlying
states.}\label{tab:msc_robustness}
\end{table}
\par\end{center}

\subsection*{Parameter Inference from Trajectories of PDK1}\label{subsec:Parameter-Inference-from}

Following analysis of calculated diffusion trajectories as decribed above, we applied our HMM parameter estimation scheme to a collection of experimental
PDK1 trajectories that were measured using single molecule TIRF microscopy, with PKC\textalpha\, concentration
near the $K_{1/2}$ for its binding to PDK1, thereby generating a population of PDK1 molecules that are bound to PKC\textalpha\ approximately half of the time.  Single molecule diffusion trajectories were imaged in movies collected at 20 msec per frame, which defines the step time.   Given the extensive computation time required for HMM analysis, the trajectories analyzed were a curated subset of representative trajectories from multiple movies collected under identical experimental conditions.  Each representative trajectory possessed long duration exceeding 375 steps, (7.5 sec total) and displayed subjective evidence for multiple diffusive modes (as in \ref{fig:sample_mo7_1b}).  Each trajectory is labeled with a number (Trajectory ``N'') identifying it 
within the collection of trajectories.

Similar to our method in Section \ref{subsec:Ensemble-Robustness-Results-1},
we performed all analyses with no prior information, formal or otherwise.
Our 1-state (Brownian motion) analysis was performed using the analytic
estimator of Equation \ref{eqn:1draylmle}. Our 2-state and 3-state HMM analyses were performed independently of one another; completing an
analysis of one did not inform our approach to the other. In practice,
sequential analysis is preferable, so that interpretation of the simpler mixture
model may inform the analysis of the hidden Markov Model. Specifically, the diffusion coefficients inferred from the mixture model may be used to constrain the parameter space of the HMM.

\subsubsection{2-State Analysis\label{sec:pdk1_2state}}

A membrane-associated, PDK1 monomer is tightly bound to its target lipid on a timescale up to multiple seconds, and its 2-dimensional diffusion on the membrane exhibits a D value defined by the frictional drag of the protein-lipid complex against the viscous bilayer.  PDK1 can reversibly associate with membrane-bound PKC\textalpha, yielding a heterodimeric complex with a much larger frictional drag.
Thus the diffusion coefficient of PDK1 in the heterodimeric complex
will be significantly smaller than that of the PDK1 monomer. Given this understanding
of a 2-state  system, we interpret the smaller diffusion coefficient,
$D_{1}$ as that of the PDK1-PKC\textalpha\ heterodimer, and the larger
diffusion coefficient as that of the PDK1 monomer. The association and dissociation transition probabilities
$p_{12}$ are thus interpreted as (a) the probability
of PDK1 and PKC\textalpha\ to form the heterodimer between consecutive frames, and (b) $p_{21}$ the probability of the PDK1-PKC\textalpha\ heterodimer to dissociate
between consecutive frames.

\begin{figure}[!htbp]
\centering{}\includegraphics[scale=0.4]{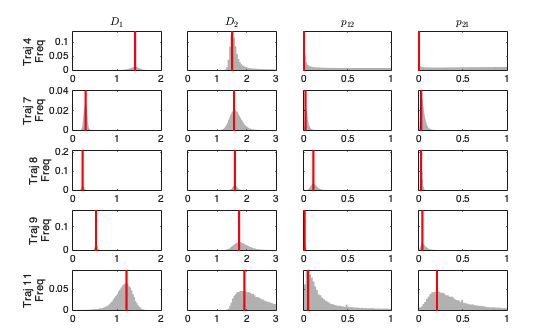}
\caption{Posterior distributions from a 2-state HMM analysis of PDK1 trajectories.}
\label{fig:pdk1_2state_posteriors}
\end{figure}

\begin{figure}[!htbp]
\centering{}\includegraphics[width=0.45\textwidth]{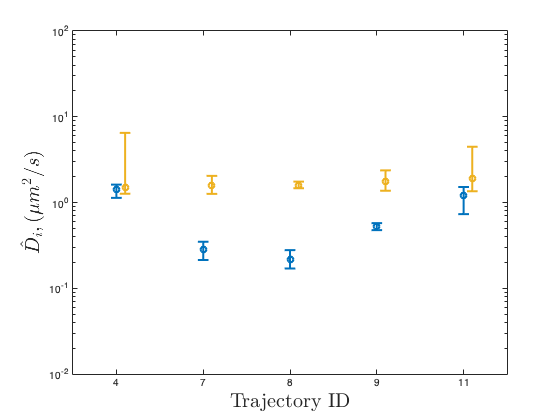}
\includegraphics[width=0.45\textwidth]{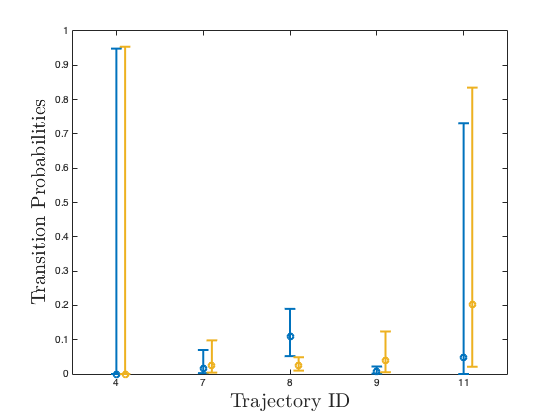}
\caption[95\% credible intervals of 2-state HMM parameters]{95\% credible intervals of 2-state HMM parameters. MLEs are indicated
by a circle, 95\% credible intervals are indicated by vertical bars.
Left: Diffusion coefficients, $D_{1}$ colored blue, $D_{2}$ colored
gold. Right: Transition probabilities, $p_{12}$ colored blue, $p_{21}$
colored gold.}
\label{fig:hmm2_bars}
\end{figure}

Figure \ref{fig:pdk1_2state_posteriors} depicts posterior distributions
of the parameters for a 2-state HMM. MLEs and credible intervals are tabulated in the Supplemental Information (Section \ref{sec:hmm2_pdk1_tables}). The marginalized posterior
distributions indicate good agreement of the value of the high-mobility diffusion coefficient, $D_2$ (ranging from about $1.5-1.9\mu\text{m}\textsuperscript{2}/\text{s}$) and, to a lesser extent, the inferred transition probabilities.
The $D_1$ posterior distributions of trajectories 4 and 11 indicate that this state is characterized by a larger
diffusion coefficient than in the other trajectories, and, uniquely,
the 95\% credible intervals of $D_{1}$ and $D_{2}$ overlap (Figure \ref{fig:hmm2_bars}). This hints at the possibility that
the states $D_{1}$ and $D_{2}$ appear in this model as two degenerate
characterizations of the same, highly diffusive state. This contrasts with the other trajectories, which exhibit both a high diffusivity state and a distinct, well-defined low diffusivity state.  The simplest explanation is that Trajectories 4 and 11 track a defective protein possessing a native lipid-binding domain that binds target lipid normally, and a non-native kinase domain unable to bind PKC\textalpha.  As a result, the proteins appear to spend the full trajectory duration in the high-diffusivity, monomeric state. In any population of proteins, at least a few defective molecules will be present, and our analysis hints this may be the case for the PDK1 molecules observed in Trajectories 4 and 11. For the three other trajectories the lower-mobility state is characterized by a diffusion coefficient $D_{1}$ in the range of $0.2-0.5\mu\text{m}\textsuperscript{2}/\text{s}$.

Figure \ref{fig:hmm2_bars} also illustrates the credibility of inferred
transition probabilities. Of note is that (excepting trajectory 11),
the per-frame probabilities of leaving either state are confidently
inferred to be significantly less than 1, which validates \textit{post hoc} that the protein generally remains in a given state at least for one full frame (20 msec), and that multiple, undetected transitions between the two observed states are unlikely to occur in a single frame.   If the state transition probabilities were larger than, $\sim0.5$, this would indicate a
violation of the assumption that state transitions are generally slower than the frame rate and thus detectable. In such a case, the frame rate would need to be increased to detect transitions.

The transition probabilities $p_{12}$ range from $0-0.11$ and $p_{21}$ range from $0-0.20$. The posterior distributions of first-order rate constants take the same shape as those for the transition probabilities (with the horizontal axis scaled according to Equation \ref{eqn:prob_to_rate}) There is strong agreement of $k_{21}$ among trajectories exhibiting evidence of two-state switching (i.e., trajectories 7, 8 and 9) yielding a narrow range of $1.5-1.9\text{s}^{-1}$. For these three trajectories, estimates of $k_{12}$ range between $0.35$ and  $5.4 \text{s}^{-1}$.

\subsubsection{3-State Analysis}

\label{sec:pdk1_3state} Recent work has uncovered previously unknown
diffusion states of PKC\textalpha\, where the major membrane-bound state is found to have two substates with the PKC\textalpha\ C1A domain inserted into the membrane in either a shallow or deep configuration. When PKC\textalpha\ is bound to PDK1, the deep configuration of C1A is predicted to confer a greater frictional drag on the complex than its shallow conformation \cite{ziemba_falke_2014}. For this analysis,
we interpret the smallest diffusion coefficient, $D_{1}$, to characterize
the diffusion of PDK1 associated with PKC\textalpha\,(deep); $D_{2}$ to characterize
the diffusion of PDK1 associated with PKC\textalpha\,(shallow); and the largest
diffusion coefficient $D_{3}$ to characterize the 
diffusion of monomeric PDK1. The transition
probabilities are interpreted accordingly, as in the 2-state analysis.
We make no assumption restricting the possibility of interchange between
any two states (for instance, the complex PDK1-PKC(deep) may transition directly
(and reversibly) to monomeric PDK1).

\begin{figure}[!htbp]
\centering \includegraphics[height=5cm]{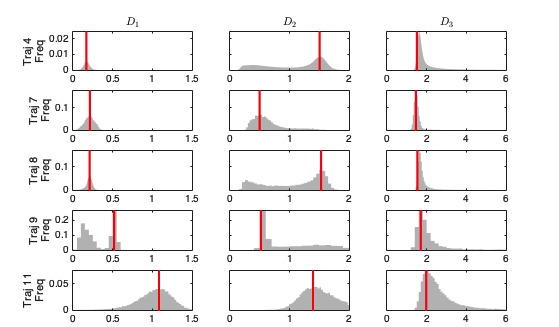}\quad{}\includegraphics[height=5cm]{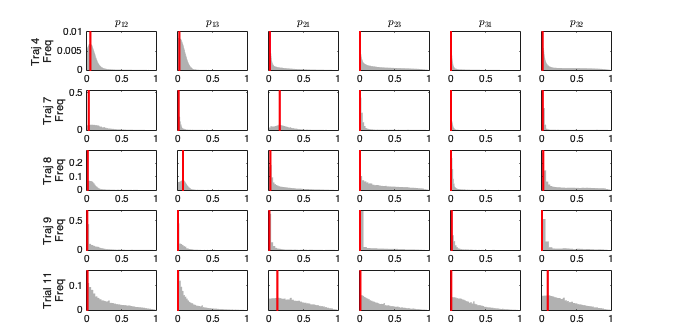}\caption[3-state HMM posterior distributions from PDK1 trajectories.]{Posterior distributions from 3-state HMM analysis of PDK1 trajectories.}
\label{fig:pdk1_hmm3_posteriors}
\end{figure}

\begin{figure}[!htbp]
\centering{}\includegraphics[height=4cm]{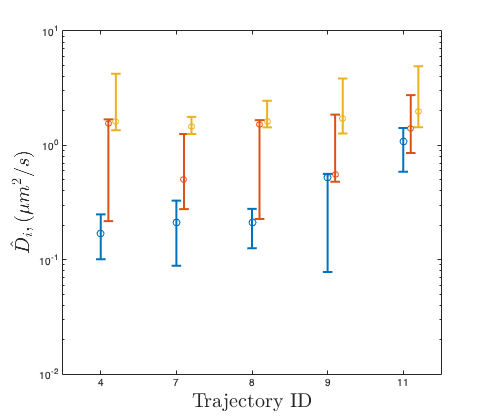}\quad{}\includegraphics[height=4cm]{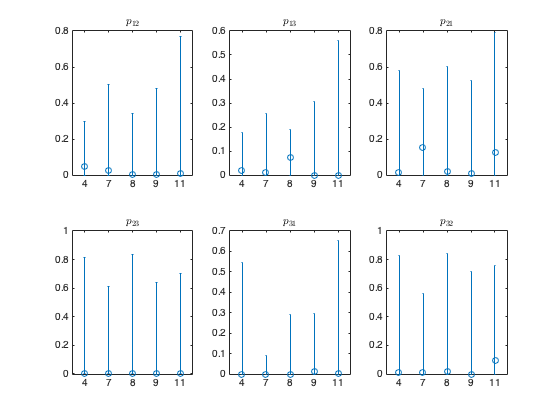}
\caption[3-state HMM credible intervals from PDK1 trajectories.]{95\% credible intervals for 3-state HMM. MLEs are indicated by a circle,
95\% credible intervals are indicated by vertical bars. Left: Diffusion
coefficients, $D_{1}$ colored blue, $D_{2}$ colored red, $D_{3}$
colored gold. Right: Transition probabilities.}
\label{fig:pdk1_hmm3_bars}
\end{figure}

Figure \ref{fig:pdk1_hmm3_posteriors} depicts posterior distributions
of the parameters for a 3-state HMM. The posterior distributions of $D_{2}$ obtained from trajectories 4 and 8
are notably bimodal, with modes around  0.25
and 1.5  $\mu\text{m}^{2}/s$ that approximately agree with
the Trajectory 8 estimates of $D_{1}$ and $D_{2}$ from the 2-state model ($0.218$
and $1.596\mu\text{m}^{2}/s$, respectively). This can be interpreted
as an identification error, where the label of $D_{2}$ can alternately
be assigned to the states characterized by $D_{1}$ and $D_{3}$.
Such a posterior distribution is suggestive that a 3-state model over-fits
the data. Interestingly, this additional degree of freedom seems to have identified a low-mobility state in Trajectory 4 that was not detected in the 2-state analysis.  Similar identification errors appear to occur in
all trajectories, as illustrated in Figure \ref{fig:pdk1_hmm3_bars}.
Specifically, the distributions of $D_{2}$ overlap significantly
with those of $D_{1}$ and $D_{ 3}$, suggesting that the states are not all well-defined and distinct.

The preference for a 2-state model over a 3-state model is supported
by both the AIC and BIC for nearly all trajectories considered, which
would seem to contradict the understood mechanism of PKC\textalpha\,activation.  The simplest explanation is that the shallow and deep states of the PDK1-PKC\textalpha\ heterodimer possess diffusivities that are too similar to be resolved by the present analysis, such that only the monomer and heterodimer states can be distinguished.
A possible exception is trajectory 7, for which the BIC indicates no
preference between a 2- and 3-state model. Interestingly, the maximum
likelihood estimates of the 3-state model for this trajectory, $D_{1}=0.2126\mu\text{m}^{2}/s$,
$D_{2}=0.4989\mu\text{m}^{2}/s$ and $D_{3}=1.4591\mu\text{m}^{2}/s$,
nearly match the corresponding measurements in \cite{ziemba_falke_2014}.
Given this agreement with a previous study, it is possible that the rate constants
tabulated in Table \ref{tab:pdk1_3state_krates} for trajectory 7
may provide preliminary insights into the transition kinetics of the full 3-state model with both shallow and deep states of the PDK1-PKC\textalpha\ heterodimer.

\begin{table}[!htbp]
\begin{centering}
\begin{tabular}{cccc}
Trajectory & 1-State & 2-State & 3-State\tabularnewline
\hline 
4 & 879 & \textbf{750} & 764\tabularnewline
7 & 577 & \textbf{327} & 336\tabularnewline
8 & 1382 & \textbf{1225} & 1244\tabularnewline
9 & 351 & \textbf{248} & 330\tabularnewline
11 & \textbf{760} & 765 & 798\tabularnewline
\end{tabular}\hfill{}%
\begin{tabular}{cccc}
Trajectory & 1-State & 2-State & 3-State\tabularnewline
\hline 
4 & 883 & \textbf{748} & 752\tabularnewline
7 & 561 & \textbf{325} & \textbf{325}\tabularnewline
8 & 1387 & \textbf{1228} & 1232\tabularnewline
9 & 355 & \textbf{246} & 318\tabularnewline
11 & 764 & \textbf{763} & 786\tabularnewline
\end{tabular}
\par\end{centering}
\caption[Model selection criteria to infer the number of states from PDK1
trajectories]{Model selection criteria. Left: AIC scores and Right: BIC scores
associated with 1, 2, 3-state HMMs for each PDK1 trajectory.}
\end{table}

\section{Conclusions and Discussion}

In this work, we presented two model classes to describe a switch
diffusion phenomena, whereby a particle's dynamics are governed by
multiple modes of Brownian motion.\footnote{All matlab software is available here: \url{https://github.com/MathBioCU/SingleMolecule}} Distinguishing these models is
their respective treatment of the unobservable sequence of diffusive
states over the course of the trajectory. In the mixture model, the
state sequence is a realization a Bernoulli scheme, where subsequent
states are effectively determined by the flip of a weighted coin.
For our HMM, the state sequence is a realization of a Markov chain\footnote{Interestingly, subsequent states are also determined by a coin flip,
but the particular weights of the coin depend on the present state.}. We developed $k$-state generalizations of both of these models,
including formulations of the likelihood functions for the purpose
of parameter inference. Furthermore, we presented detailed kinetic
interpretations of the model parameters, contending that the HMM provides
a more valuable description of kinetics of membrane-targeting proteins.

We build upon previous work \cite{coombs2009}, which prescribes an
MCMC scheme to parameter inference of a two-state HMM, by extending
their methodology to consider both a 3-state HMM and mixture model.
This extension uncovered several nontrivial practical challenges that
make an automated analysis computationally and analytically difficult.
Nevertheless, we improved upon the work of Das to develop a ``black
box'' algorithm to evaluate the maximum likelihood estimators of
a model for a given trajectory, with minimal user-specified tuning.
We achieved this by implementing an affine invariant ensemble MCMC
sampler (GWMCMC), which efficiently samples posterior distributions
for badly-scaled problems.

We empirically assessed the robustness of the method by testing it
on an ensemble of simulated trajectories, which were constructed from
a very general set of underlying parameters. One component of our
robustness study is a comparison between inference of parameters between
mixture models and HMMs. Our results indicate, unsurprisingly, that
the mixture model is more robust and tends to converge more quickly.
Another component of this study was an evaluation of the ability of
two model selection criteria, AIC and BIC, to identify the number
of underlying diffusive states from data. Surprisingly, we found that
prediction of state size was exceptionally accurate for both model
selection criteria, across both model classes (Table \ref{tab:msc_robustness}), and that that these state predictions obtained from HMMs were more reliable, even though parameter inference appears more robust for mixture models (Figure \ref{fig:relative_error_plots} and Tables \ref{tab:mr2_bias_summary}-\ref{tab:hmm3_bias_summary}).

The robustness test that we use in the present work was designed to evaluate the ability of the method to infer parameters under an extremely broad range of conditions. Our analysis indicates that the inference scheme typically performs quite well, even as the number of diffusive states increases. In practice, one may reasonably expect the parameter estimation schemes to perform better than indicated by the robustness test, since the range of parameters chosen for the robustness analysis are more broad than the parameters would be in their expected use case. Specifically, the HMM analysis studied here is best suited for systems where the states exhibit long dwell times, since an assumption of the physical system is that state transitions occur on a long timescale (long dwell time) relative to the time span of an individual movie frame or diffusion step. Our robustness study included trajectories with very short dwell times, with the intention of assessing the efficacy of the method under the widest practicable range of conditions. A robustness study that more accurately reflects the experimental conditions that give rise to SPT trajectories (e.g. restricting $p_{ij}\ll1$ for $i\neq j$, or requiring that the HMM transition matrix be diagonally dominant) may be a more representative study. Our study, however, indicates that the long dwell time assumption can be validated (or rejected) \textit{post hoc} with reasonably high accuracy.

One can implement formal priors
to reflect an understanding of the system. A significant advantage
to a Bayesian approach to parameter inference is that priors may be
used to reflect one's beliefs about the parameters giving rise to
a stochastic process. We endeavored here to produce a robust approach
to parameter inference that required as little \textit{a priori} information
as possible. An expression of this agnostic approach was the disuse
of prior distributions. A specific choice of a prior that might be
selected is a beta distribution (or multivariate beta distribution,
as appropriate) for the transition probabilities of the HMM, to reflect
the prediction that transition probabilities will be small in magnitude
compared to 1.

Efforts to reduce the dimension of parameter space are demonstrably
worthwhile to improve the robustness of inference, or to at least
decrease computational cost. To improve inference of transition probabilities
and rate constants, one may wish to use well-established literature diffusion coefficients, or compute diffusion coefficients using a mixture model, then use these predeterminned D values to initialize the  sampler to decrease
the computational time and result in higher fidelity sampling of modeled diffusion coefficients.  Or, such predetermined diffusion coefficients can be supplied  to the hidden Markov model to substantially reduce the dimensionality and computational time of its search.  

We applied our method to predict rate constants for interactions
of PDK1 with PKC\textalpha. We presented a detailed interpretation of the posterior distributions of our model parameters. Our analysis highlights the challenge of parameter identifiability in this context. Although the literature \cite{Gordon2021} and data including Figure 1b suggest the existence of three diffusion states for this system, our HMM analysis provides a more stable fit to the data for a two-state model than a three-state model. Initializing the GWMCMC walkers with the three-state diffusion coefficients obtained from a previous study dramatically reduced number of iterations necessary for convergence, but did not produce a significantly better fit. Likewise, fixing the diffusion coefficients using those obtained from a previous study resulted in further acceleration of convergence, but did not result in less uncertainty in the extracted transition probabilities. Notably, the three literature diffusion coefficients employed include one distinct value ($D = 1.7 \mu$m\protect\protect\textsuperscript{2}/s corresponding to the high diffusivity PDK1 monomer state) and two smaller values similar in magnitude ($D = 0.41\mu$m\protect\protect\textsuperscript{2}/s and $D = 0.17\mu$m\protect\protect\textsuperscript{2}/s corresponding to the PDK1-PKC\textalpha\   heterodimer with PKC\textalpha\ in its shallow and deep states). The simplest possibility is that the present approach is unable to resolve the latter two heterodimer states. It follows that full resolution of three diffusion states would only be possible if their diffusion coefficients were more distinct, or with a larger input data set.  We did find adequate agreement to report reasonable estimates for the first order transition probabilities of a two-state model in which PDK1 switches between a kinetically stable monomeric state and a kinetically stable PDK1, PKC\textalpha\ heterodimer.  This two-state model is sufficient to describe the kinetics of the monomer-heterodimer interconversion and provides, to our knowledge, the first kinetic scheme for this biologically crucial signaling reaction occurring on a membrane surface.

Although the findings of the two-state HMM analysis are better suited for elucidating the basic kinetics of switching between simple, kinetically stable monomer and heterodimer states, we hypothesize that the three-state HMM analysis may reveal an additional complexity present in the monomer-heterodimer system.   Specifically, in the three-state analysis, the middle, broad diffusivity state (D2) is observed to exhibit a range of diffusion constants that can encompass both monomer and heterodimer diffusivities.  We hypothesize this D2 state could represent an unstable rapid interconversion state wherein PDK has a PKC nearby and is rapidly switching between monomer (D3) and heterodimer (D1) states.  When no PKC is nearby, the kinetically stable monomer state (D3) is observed.  When PDK and PKC combine with optimal protein-protein and protein-membrane contacts, the kinetically stable heterodimer state (D1) is observed.  In the rapid interconversion state (D2) the optimal protein and membrane contacts are not yet fully formed, leading to the rapid switching between the unbound (monomer) and bound (heterodimer) states.  Notably, the rapid interconversion state (D2) would represent a novel, previously unobserved intermediate state in the association reaction between two proteins on a membrane surface.  Moreover, we propose the three HMM states D1, D2 and D3 may correspond to the slow, intermediate and fast states typically observed in running averages of total displacement per frame vs time, as illustrated in Fig 1B above.  In short, this interpretation of the three-state HMM model still needs to be directly tested, but offers an exciting direction for future study.

\section{Acknowledgments}
This research was supported in part by the NIH grants  R01GM063235 and R35GM144346 (to JJF), T32GM065103 (to MTG), R01GM126559 (to DMB) and NSF grant 2054085 (to DMB). This work also utilized resources from the University of Colorado Boulder Research Computing Group, which is supported by the National Science Foundation (awards ACI-1532235 and ACI-1532236), the University of Colorado Boulder, and Colorado State University. The authors would also like to thank Vanja Duki\'c (Department of Applied Mathematics, University of Colorado) for helpful discussions concerning the Bayesian statistics and the sampling strategy.

\bibstyle{plos2015}
\nocite{*}
\bibliography{masterbib}

\begin{thebibliography}{10}

\bibitem{coombs2009}
Das R, Cairo CW, Coombs D.
\newblock A hidden Markov model for single particle tracks quantifies dynamic
  interactions between LFA-1 and the actin cytoskeleton.
\newblock PLoS Comput Biol. 2009;5(11):e1000556.

\bibitem{BanksBortz2005JInverseIll-PosedProbl}
Banks HT, Bortz DM.
\newblock Inverse Problems for a Class of Measure Dependent Dynamical Systems.
\newblock J Inverse Ill-Posed Probl. 2005;13(2):103--121.
\newblock doi:{10.1515/1569394053978515}.

\bibitem{MirzaevByrneBortz2016InverseProbl}
Mirzaev I, Byrne EC, Bortz DM.
\newblock An {{Inverse Problem}} for a {{Class}} of {{Conditional Probability
  Measure-Dependent Evolution Equations}}.
\newblock Inverse Probl. 2016;32(9):095005.
\newblock doi:{10.1088/0266-5611/32/9/095005}.

\bibitem{knight2009single}
Knight JD, Falke JJ.
\newblock Single-molecule fluorescence studies of a PH domain: new insights
  into the membrane docking reaction.
\newblock Biophysical journal. 2009;96(2):566--582.

\bibitem{falke2010}
Knight JD, Lerner MG, Marcano-Vel{\'a}zquez JG, Pastor RW, Falke JJ.
\newblock Single molecule diffusion of membrane-bound proteins: window into
  lipid contacts and bilayer dynamics.
\newblock Biophysical journal. 2010;99(9):2879--2887.

\bibitem{ziemba2013lateral}
Ziemba BP, Falke JJ.
\newblock Lateral diffusion of peripheral membrane proteins on supported lipid
  bilayers is controlled by the additive frictional drags of (1) bound lipids
  and (2) protein domains penetrating into the bilayer hydrocarbon core.
\newblock Chemistry and physics of lipids. 2013;172:67--77.

\bibitem{ZiembaKnightFalke2012Biochemistry}
Ziemba BP, Knight JD, Falke JJ.
\newblock Assembly of {{Membrane-Bound Protein Complexes}}: {{Detection}} and
  {{Analysis}} by {{Single Molecule Diffusion}}.
\newblock Biochemistry. 2012;51(8):1638--1647.
\newblock doi:{10.1021/bi201743a}.

\bibitem{BucklesZiembaMassonEtAl2017BiophysicalJournal}
Buckles TC, Ziemba BP, Masson GR, Williams RL, Falke JJ.
\newblock Single-{{Molecule Study Reveals How Receptor}} and {{Ras
  Synergistically Activate PI3K$\alpha$}} and {{PIP3 Signaling}}.
\newblock Biophysical Journal. 2017;113(11):2396--2405.
\newblock doi:{10.1016/j.bpj.2017.09.018}.

\bibitem{BucklesOhashiTremelEtAl2020BiophysicalJournal}
Buckles TC, Ohashi Y, Tremel S, McLaughlin SH, Pardon E, Steyaert J, et~al.
\newblock The {{G-Protein Rab5A Activates VPS34 Complex II}}, a {{Class III
  PI3K}}, by a {{Dual Regulatory Mechanism}}.
\newblock Biophysical Journal. 2020;119(11):2205--2218.
\newblock doi:{10.1016/j.bpj.2020.10.028}.

\bibitem{Gordon2021}
Gordon MT, Ziemba BP, Falke JJ.
\newblock Single Molecule Studies and Kinase Activity Measurements Reveal
  Regulatory Interactions between the Master Kinases
  Phosphoinositide-Dependent-Kinase-1 (PDK1), Protein Kinase B (AKT1/PKB) and
  Protein Kinase C (PKC$\alpha$).
\newblock Biophysical Journal. 2021;120(24).

\bibitem{kusumi1993confined}
Kusumi A, Sako Y, Yamamoto M.
\newblock Confined lateral diffusion of membrane receptors as studied by single
  particle tracking (nanovid microscopy). Effects of calcium-induced
  differentiation in cultured epithelial cells.
\newblock Biophysical journal. 1993;65(5):2021--2040.

\bibitem{slator2018}
Slator PJ, Burroughs NJ.
\newblock A hidden Markov model for detecting confinement in single-particle
  tracking trajectories.
\newblock Biophysical journal. 2018;115(9):1741--1754.

\bibitem{schutz1997single}
Sch{\"u}tz GJ, Schindler H, Schmidt T.
\newblock Single-molecule microscopy on model membranes reveals anomalous
  diffusion.
\newblock Biophysical journal. 1997;73(2):1073--1080.

\bibitem{yogurtcu2018cytosolic}
Yogurtcu ON, Johnson ME.
\newblock Cytosolic proteins can exploit membrane localization to trigger
  functional assembly.
\newblock PLoS computational biology. 2018;14(3):e1006031.

\bibitem{ziemba_falke_2014}
Ziemba BP, Li J, Landgraf KE, Knight JD, Voth GA, Falke JJ.
\newblock Single-molecule studies reveal a hidden key step in the activation
  mechanism of membrane-bound protein kinase C-$\alpha$.
\newblock Biochemistry. 2014;53(10):1697--1713.

\bibitem{saxton1997single}
Saxton MJ.
\newblock Single-particle tracking: the distribution of diffusion coefficients.
\newblock Biophysical journal. 1997;72(4):1744--1753.

\bibitem{slator2015}
Slator PJ, Cairo CW, Burroughs NJ.
\newblock Detection of diffusion heterogeneity in single particle tracking
  trajectories using a hidden Markov model with measurement noise propagation.
\newblock PloS one. 2015;10(10):e0140759.

\bibitem{qian1991single}
Qian H, Sheetz MP, Elson EL.
\newblock Single particle tracking. Analysis of diffusion and flow in
  two-dimensional systems.
\newblock Biophysical journal. 1991;60(4):910--921.

\bibitem{dutil1998regulation}
Dutil EM, Toker A, Newton AC.
\newblock Regulation of conventional protein kinase C isozymes by
  phosphoinositide-dependent kinase 1 (PDK-1).
\newblock Current Biology. 1998;8(25):1366--1375.

\bibitem{leonard2011crystal}
Leonard TA, R{\'o}{\. z}ycki B, Saidi LF, Hummer G, Hurley JH.
\newblock Crystal structure and allosteric activation of protein kinase C
  $\beta$II.
\newblock Cell. 2011;144(1):55--66.

\bibitem{koo2015extracting}
Koo PK, Weitzman M, Sabanaygam CR, van Golen KL, Mochrie SG.
\newblock Extracting diffusive states of Rho GTPase in live cells: towards in
  vivo biochemistry.
\newblock PLoS computational biology. 2015;11(10):e1004297.

\bibitem{linden2018variational}
Lind{\'e}n M, Elf J.
\newblock Variational algorithms for analyzing noisy multistate diffusion
  trajectories.
\newblock Biophysical journal. 2018;115(2):276--282.

\bibitem{elf2019single}
Elf J, Barkefors I.
\newblock Single-molecule kinetics in living cells.
\newblock Annual review of biochemistry. 2019;88:635--659.

\bibitem{falcao2020diffusion}
Falcao RC, Coombs D.
\newblock Diffusion analysis of single particle trajectories in a Bayesian
  nonparametrics framework.
\newblock Physical biology. 2020;17(2):025001.

\bibitem{geissen2019inference}
Geissen EM, Hasenauer J, Radde NE.
\newblock Inference of finite mixture models and the effect of binning.
\newblock Statistical applications in genetics and molecular biology.
  2019;18(4).

\bibitem{bullerjahn2021maximum}
Bullerjahn JT, Hummer G.
\newblock Maximum likelihood estimates of diffusion coefficients from
  single-particle tracking experiments.
\newblock The Journal of Chemical Physics. 2021;154(23):234105.

\bibitem{sitrin2010migrating}
Sitrin RG, Sassanella TM, Landers JJ, Petty HR.
\newblock Migrating human neutrophils exhibit dynamic spatiotemporal variation
  in membrane lipid organization.
\newblock American journal of respiratory cell and molecular biology.
  2010;43(4):498--506.

\bibitem{persson2013extracting}
Persson F, Lind{\'e}n M, Unoson C, Elf J.
\newblock Extracting intracellular diffusive states and transition rates from
  single-molecule tracking data.
\newblock Nature methods. 2013;10(3):265--269.

\bibitem{pohle2017selecting}
Pohle J, Langrock R, van Beest FM, Schmidt NM.
\newblock Selecting the number of states in hidden Markov models: pragmatic
  solutions illustrated using animal movement.
\newblock Journal of Agricultural, Biological and Environmental Statistics.
  2017;22(3):270--293.

\bibitem{bp1969}
Petrie T.
\newblock Probabilistic functions of finite state Markov chains.
\newblock The Annals of Mathematical Statistics. 1969;40(1):97--115.

\bibitem{gwmcmc}
Goodman J, Weare J.
\newblock Ensemble samplers with affine invariance.
\newblock Communications in applied mathematics and computational science.
  2010;5(1):65--80.

\bibitem{emcee}
Foreman-Mackey D, Hogg DW, Lang D, Goodman J.
\newblock emcee: the MCMC hammer.
\newblock Publications of the Astronomical Society of the Pacific.
  2013;125(925):306.

\bibitem{sokal1996}
Sokal A.
\newblock Monte Carlo methods in statistical mechanics: foundations and new
  algorithms.
\newblock In: Functional integration. Springer; 1997. p. 131--192.

\bibitem{schwarz1978}
Schwarz G, et~al.
\newblock Estimating the dimension of a model.
\newblock Annals of statistics. 1978;6(2):461--464.

\bibitem{saxton2008single}
Saxton MJ.
\newblock Single-particle tracking: connecting the dots.
\newblock Nature methods. 2008;5(8):671--672.

\bibitem{manzo2015review}
Manzo C, Garcia-Parajo MF.
\newblock A review of progress in single particle tracking: from methods to
  biophysical insights.
\newblock Reports on progress in physics. 2015;78(12):124601.

\bibitem{casellaberger2002}
Casella G, Berger RL.
\newblock Statistical inference.
\newblock Cengage Learning; 2002.

\bibitem{Leroux_1990}
Leroux BG.
\newblock Maximum-likelihood estimation for hidden Markov models.
\newblock Stochastic Processes and their Applications. 1992;40(1):127--143.
\newblock doi:{10.1016/0304-4149(92)90141-c}.

\bibitem{gilks1996}
Gilks R, Richardson S.
\newblock Markov Chain Monte Carlo in practice.
\newblock Chapman and Hall/CRC; 1996.

\bibitem{burnham2002}
Burnham KP, Anderson DR.
\newblock Model selection and multimodel inference. vol.~2.
\newblock Springer New York; 2002.

\bibitem{press2007}
Press WH, William H, Teukolsky SA, Vetterling WT, Saul A, Flannery BP.
\newblock Numerical recipes 3rd edition: The art of scientific computing.
\newblock Cambridge university press; 2007.

\bibitem{fish2009total}
Fish KN.
\newblock Total internal reflection fluorescence (TIRF) microscopy.
\newblock Current protocols in cytometry. 2009;50(1):12--18.

\bibitem{bacallado2009bayesian}
Bacallado S, Chodera JD, Pande V.
\newblock Bayesian comparison of Markov models of molecular dynamics with
  detailed balance constraint.
\newblock The Journal of chemical physics. 2009;131(4):07B622.

\bibitem{deleeuw1992introduction}
deLeeuw J.
\newblock Introduction to Akaike (1973) Information theory and an extension of
  the maximum likelihood principle.
\newblock In: Breakthroughs in statistics. Springer; 1992. p. 599--609.

\bibitem{grinsted_gwmcmc}
Grinsted A.
\newblock Ensemble MCMC sampler (https://github.com/grinsted/gwmcmc), GitHub.
  Retrieved March 9, 2018.
\newblock MATLAB Central. 2018;.

\bibitem{saxton1990lateral}
Saxton MJ.
\newblock Lateral diffusion in a mixture of mobile and immobile particles. A
  Monte Carlo study.
\newblock Biophysical journal. 1990;58(5):1303--1306.

\end{thebibliography}

\appendix

\section{Appendix}

\subsection{One Diffusive State\label{subsec:One-Diffusive-State}}

We'll begin with a brief discussion of 2-D Brownian diffusion. A particle
subject to 2-D Brownian diffusion whose location is sampled $N$ times
after fixed time intervals $\tau$ will produce a sequence of independent
displacements $\mathbf{O}=(\mathbf{r}_{i})_{i=1}^{N}$ with $\mathbf{r}_{i}=(x_{i},y_{i})$.
In simple Brownian diffusion, the coordinate displacements $x_{i}$
and $y_{i}$ are independent and Normally distributed with mean 0
and variance $2D\tau$. These coordinate displacements are typically
synthesized into a net displacement $r_{i}=\vert\mathbf{r}_{i}\vert=\sqrt{x_{i}^{2}+y_{i}^{2}}$.
It's straightforward to show that $r_{i}$ is Rayleigh-distributed
(a rescaling of the Chi distribution with two degrees of freedom),
with pdf 
\begin{equation}
f(r|D,\tau)=\frac{1}{4D\tau}e^{-r^{2}/(4D\tau)}.\label{eqn:1drayl}
\end{equation}
The likelihood of a given diffusion coefficient $D$, given an observation
$\mathbf{O}$ is given by 
\[
L(D\vert\mathbf{O})\propto\prod_{i=1}^{N}f(r_{i}|D,\tau)=\frac{1}{4D\tau}\exp\left(-\frac{1}{4D\tau}\sum_{i=1}^{N}r_{i}\right).
\]
The maximum likelihood estimator of the diffusion coefficient $D_{\text{MLE}}$
is the value of $D$ that maximizes this quantity or, equivalently,
the log likelihood 
\[
\ell(D\vert\mathbf{O})=\log\left(L(D\vert\mathbf{O}\right)=-N\log(D\tau)-\frac{1}{4D\tau}\sum_{i=1}^{N}r_{i}^{2}.
\]
This function of $D$ is maximized for 
\begin{equation}
D_{\text{MLE}}=\frac{1}{4\tau N}\sum_{i=1}^{N}r_{i}^{2}.\label{eqn:1draylmle}
\end{equation}
In other words, the diffusion coefficient that is best supported an
observation $\mathbf{O}$ is proportional to the average squared displacement.
This method of analysis is typical in the biochemistry literature
and produces a formula familiar to biochemists who study diffusion,
\[
\text{ssd}(T)=\sum_{i=1}^{N}r_{i}^{2}=4DT
\]
where $\text{ssd}(T)$ is the sum of squared displacements observed
over a trajectory of length $T=N\tau$.


\section{Supplemental Information}

\subsection{MCMC Particulars}
Parameter inference was performed using affine-invariant Markov Chain Monte Carlo (GWMCMC). Initial transition probabilities and mixing coefficients were initialized randomly, in accordance with the law of total probability. Diffusion coefficients were likewise initialized randomly from $\text{Unif}[D_{\min},D_{\max}]$. In all results presented here, GWMCMC was performed with 200 walkers. Chain convergence was checked every $10^6$ steps.

\begin{center}
\begin{figure}
\centering{}\includegraphics[width=0.45\textwidth]{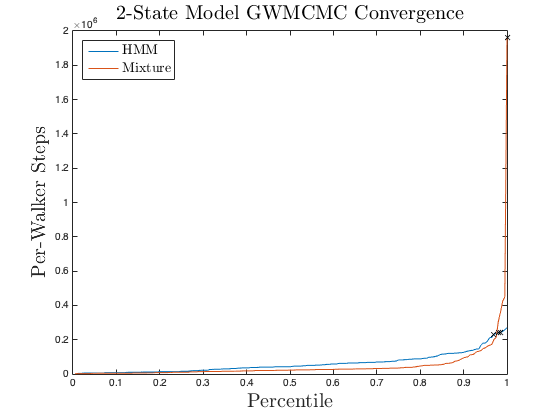}
\includegraphics[width=0.45\textwidth]{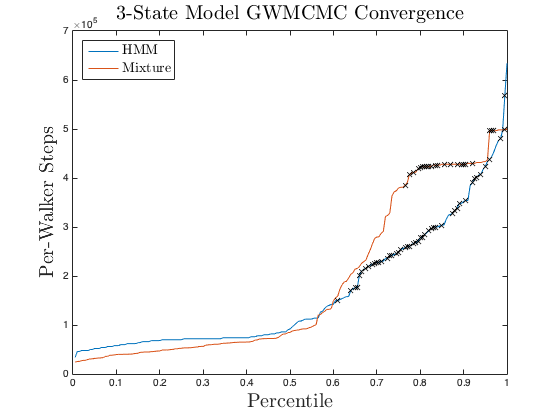}
\caption[Summary of ensemble MCMC chain lengths]{Length of GWMCMC chains, per walker. Left: 3-state models. Right:
2-state models. Note the different in vertical scaling. Trials that
did not achieve convergence are indicated by a black 'x'.}
\label{fig:mcmc_convergence_plots}
\end{figure}
\par\end{center}

\subsection{Results of HMM Analysis of PDK1 Trajectories}

\subsubsection{Tabulated 2-State HMM Results\label{sec:hmm2_pdk1_tables}}
\begin{center}
\begin{table}[!htb]
{\tiny{}}\subfloat[Diffusion coefficients and transition probabilities.]{\begin{centering}
{\tiny{}}%
\begin{tabular}{ccr@{\extracolsep{0pt}.}lr@{\extracolsep{0pt}.}lr@{\extracolsep{0pt}.}lr@{\extracolsep{0pt}.}lr@{\extracolsep{0pt}.}lr@{\extracolsep{0pt}.}lr@{\extracolsep{0pt}.}lr@{\extracolsep{0pt}.}lr@{\extracolsep{0pt}.}lr@{\extracolsep{0pt}.}lr@{\extracolsep{0pt}.}lr@{\extracolsep{0pt}.}l}
 & {\tiny{}Steps} & \multicolumn{2}{c}{{\tiny{}$\hat{D}_{1}$}} & \multicolumn{2}{c}{} & \multicolumn{2}{c}{} & \multicolumn{2}{c}{{\tiny{}$\hat{D}_{2}$}} & \multicolumn{2}{c}{} & \multicolumn{2}{c}{} & \multicolumn{2}{c}{{\tiny{}$\hat{p}_{12}$}} & \multicolumn{2}{c}{} & \multicolumn{2}{c}{} & \multicolumn{2}{c}{{\tiny{}$\hat{p}_{21}$}} & \multicolumn{2}{c}{} & \multicolumn{2}{c}{}\tabularnewline
\hline 
{\tiny{}Traj 4} & {\tiny{}448} & {\tiny{}1}&{\tiny{}4126} & {\tiny{}1}&{\tiny{}1304} & {\tiny{}1}&{\tiny{}6168} & {\tiny{}1}&{\tiny{}5024} & {\tiny{}1}&{\tiny{}2636} & {\tiny{}6}&{\tiny{}4763} & {\tiny{}0}&{\tiny{}0} & {\tiny{}0}&{\tiny{}0} & {\tiny{}0}&{\tiny{}9487} & {\tiny{}0}&{\tiny{}0} & {\tiny{}0}&{\tiny{}0} & {\tiny{}0}&{\tiny{}9540}\tabularnewline
{\tiny{}Traj 7} & {\tiny{}636} & {\tiny{}0}&{\tiny{}2869} & {\tiny{}0}&{\tiny{}2138} & {\tiny{}0}&{\tiny{}3490} & {\tiny{}1}&{\tiny{}5690} & {\tiny{}1}&{\tiny{}2621} & {\tiny{}2}&{\tiny{}0464} & {\tiny{}0}&{\tiny{}0173} & {\tiny{}0}&{\tiny{}0027} & {\tiny{}0}&{\tiny{}0700} & {\tiny{}0}&{\tiny{}0297} & {\tiny{}0}&{\tiny{}0108} & {\tiny{}0}&{\tiny{}0676}\tabularnewline
{\tiny{}Traj 8} & {\tiny{}751} & {\tiny{}0}&{\tiny{}2176} & {\tiny{}0}&{\tiny{}1698} & {\tiny{}0}&{\tiny{}2780} & {\tiny{}1}&{\tiny{}5961} & {\tiny{}1}&{\tiny{}4642} & {\tiny{}1}&{\tiny{}7513} & {\tiny{}0}&{\tiny{}1080} & {\tiny{}0}&{\tiny{}0520} & {\tiny{}0}&{\tiny{}1900} & {\tiny{}0}&{\tiny{}0239} & {\tiny{}0}&{\tiny{}0099} & {\tiny{}0}&{\tiny{}0489}\tabularnewline
{\tiny{}Traj 9} & {\tiny{}650} & {\tiny{}0}&{\tiny{}5228} & {\tiny{}0}&{\tiny{}4758} & {\tiny{}0}&{\tiny{}5753} & {\tiny{}1}&{\tiny{}7426} & {\tiny{}1}&{\tiny{}3730} & {\tiny{}2}&{\tiny{}3710} & {\tiny{}0}&{\tiny{}0070} & {\tiny{}0}&{\tiny{}0010} & {\tiny{}0}&{\tiny{}0220} & {\tiny{}0}&{\tiny{}0380} & {\tiny{}0}&{\tiny{}0050} & {\tiny{}0}&{\tiny{}1240}\tabularnewline
{\tiny{}Traj 11} & {\tiny{}375} & {\tiny{}1}&{\tiny{}2154} & {\tiny{}0}&{\tiny{}7326} & {\tiny{}1}&{\tiny{}5157} & {\tiny{}1}&{\tiny{}9203} & {\tiny{}1}&{\tiny{}3538} & {\tiny{}4}&{\tiny{}4696} & {\tiny{}0}&{\tiny{}0480} & {\tiny{}0}&{\tiny{}0} & {\tiny{}0}&{\tiny{}7310} & {\tiny{}0}&{\tiny{}2040} & {\tiny{}0}&{\tiny{}0210} & {\tiny{}0}&{\tiny{}8350}\tabularnewline
\end{tabular}{\tiny\par}
\par\end{centering}
{\tiny{}}{\tiny\par}}{\tiny\par}
\begin{centering}
{\tiny{}}\subfloat[First-order rate constants.]{\begin{centering}
{\tiny{}}%
\begin{tabular}{ccr@{\extracolsep{0pt}.}lr@{\extracolsep{0pt}.}lr@{\extracolsep{0pt}.}lr@{\extracolsep{0pt}.}lr@{\extracolsep{0pt}.}lr@{\extracolsep{0pt}.}l}
 & {\tiny{}Steps} & \multicolumn{2}{c}{{\tiny{}$\hat{k}_{12}$}} & \multicolumn{2}{c}{} & \multicolumn{2}{c}{} & \multicolumn{2}{c}{{\tiny{}$\hat{k}_{21}$}} & \multicolumn{2}{c}{} & \multicolumn{2}{c}{}\tabularnewline
\hline 
{\tiny{}Traj 4} & {\tiny{}448} & {\tiny{}0}&{\tiny{}0} & {\tiny{}0}&{\tiny{}0} & {\tiny{}47}&{\tiny{}} & {\tiny{}0}&{\tiny{}0} & {\tiny{}0}&{\tiny{}0} & {\tiny{}48}&{\tiny{}}\tabularnewline
{\tiny{}Traj 7} & {\tiny{}636} & {\tiny{}0}&{\tiny{}87} & {\tiny{}0}&{\tiny{}14} & {\tiny{}3}&{\tiny{}5} & {\tiny{}1}&{\tiny{}5} & {\tiny{}0}&{\tiny{}54} & {\tiny{}3}&{\tiny{}4}\tabularnewline
{\tiny{}Traj 8} & {\tiny{}751} & {\tiny{}5}&{\tiny{}4} & {\tiny{}2}&{\tiny{}6} & {\tiny{}9}&{\tiny{}5} & {\tiny{}1}&{\tiny{}2} & {\tiny{}0}&{\tiny{}25} & {\tiny{}6}&{\tiny{}2}\tabularnewline
{\tiny{}Traj 9} & {\tiny{}650} & {\tiny{}0}&{\tiny{}35} & {\tiny{}0}&{\tiny{}05} & {\tiny{}1}&{\tiny{}1} & {\tiny{}1}&{\tiny{}9} & {\tiny{}0}&{\tiny{}25} & {\tiny{}6}&{\tiny{}2}\tabularnewline
{\tiny{}Traj 11} & {\tiny{}375} & {\tiny{}2}&{\tiny{}4} & {\tiny{}0}&{\tiny{}0} & {\tiny{}37}&{\tiny{}} & {\tiny{}10}&{\tiny{}} & {\tiny{}1}&{\tiny{}1} & {\tiny{}42}&{\tiny{}}\tabularnewline
\end{tabular}{\tiny\par}
\par\end{centering}
{\tiny{}}{\tiny\par}}{\tiny\par}
\par\end{centering}
\centering{}{\tiny{}\caption[2-State HMM diffusion coefficient and transition probability MAPs
and 95\% credible intervals from analysis of PDK1 trajectories]{{\tiny{}\label{tab:pdk1_2state}}MLEs and associated 95\% confidence
intervals from a 2-state HMM analysis of PDK1 trajectories.}
}{\tiny\par}
\end{table}
\par\end{center}

\subsubsection{Tabulated 3-State HMM Results}

\begin{table}[!htb]
\begin{centering}
{\tiny{}}%
\begin{tabular}{ccccccccccc}
 & {\tiny{}Steps} & {\tiny{}$\hat{D}_{1}$} &  &  & {\tiny{}$\hat{D}_{2}$} &  &  & {\tiny{}$\hat{D}_{3}$} &  & \tabularnewline
\hline 
{\tiny{}Traj 4} & {\tiny{}448} & {\tiny{}0.1682} & {\tiny{}0.1009} & {\tiny{}0.2490} & {\tiny{}1.5070} & {\tiny{}0.2172} & {\tiny{}1.6878} & {\tiny{}1.6155} & {\tiny{}1.3561} & {\tiny{}4.2285}\tabularnewline
{\tiny{}Traj 7} & {\tiny{}636} & {\tiny{}0.2126} & {\tiny{}0.0884} & {\tiny{}0.3285} & {\tiny{}0.4989} & {\tiny{}0.2765} & {\tiny{}1.2554} & {\tiny{}1.4591} & {\tiny{}1.2576} & {\tiny{}1.7735}\tabularnewline
{\tiny{}Traj 8} & {\tiny{}751} & {\tiny{}0.2094} & {\tiny{}0.1257} & {\tiny{}0.2780} & {\tiny{}1.5318} & {\tiny{}0.2268} & {\tiny{}1.6620} & {\tiny{}1.6318} & {\tiny{}1.4372} & {\tiny{}2.4550}\tabularnewline
{\tiny{}Traj 9} & {\tiny{}650} & {\tiny{}0.5191} & {\tiny{}0.0781} & {\tiny{}0.5627} & {\tiny{}0.5557} & {\tiny{}0.4808} & {\tiny{}1.8561} & {\tiny{}1.7042} & {\tiny{}1.2686} & {\tiny{}3.8384}\tabularnewline
{\tiny{}Traj 11} & {\tiny{}375} & {\tiny{}1.0840} & {\tiny{}0.5868} & {\tiny{}1.4190} & {\tiny{}1.3941} & {\tiny{}0.8561} & {\tiny{}2.7489} & {\tiny{}1.9776} & {\tiny{}1.4379} & {\tiny{}4.9190}\tabularnewline
\end{tabular}{\tiny{}\label{tab:pdk1_3state_dcoeffs}}{\tiny\par}
\par\end{centering}
{\tiny{}\caption[3-State HMM diffusion coefficient MAPs and 95\% credible intervals
from analysis of PDK1 trajectories]{Diffusion coefficients estimated using a 3-state HMM obtained from
analysis of PDK1 trajectories, along with 95\% credible intervals.
Values are presented in units of $\mu\text{{m}}^{2}/\text{{s}.}$}
}{\tiny\par}
\end{table}

\begin{table}[!htb]
\centering
{\tiny{}}%
\begin{tabular}{ccr@{\extracolsep{0pt}.}lr@{\extracolsep{0pt}.}lr@{\extracolsep{0pt}.}lr@{\extracolsep{0pt}.}lr@{\extracolsep{0pt}.}lr@{\extracolsep{0pt}.}lr@{\extracolsep{0pt}.}lr@{\extracolsep{0pt}.}lr@{\extracolsep{0pt}.}l}
 & {\tiny{}Steps} & \multicolumn{2}{c}{{\tiny{}$\hat{p}_{12}$}} & \multicolumn{2}{c}{} & \multicolumn{2}{c}{} & \multicolumn{2}{c}{{\tiny{}$\hat{p}_{13}$}} & \multicolumn{2}{c}{} & \multicolumn{2}{c}{} & \multicolumn{2}{c}{{\tiny{}$\hat{p}_{21}$}} & \multicolumn{2}{c}{} & \multicolumn{2}{c}{}\tabularnewline
\hline 
{\tiny{}Traj 4} & {\tiny{}448} & {\tiny{}0}&{\tiny{}0492} & {\tiny{}0}&{\tiny{}0} & {\tiny{}0}&{\tiny{}2954} & {\tiny{}0}&{\tiny{}0199} & {\tiny{}0}&{\tiny{}0} & {\tiny{}0}&{\tiny{}1764} & {\tiny{}0}&{\tiny{}0120} & {\tiny{}0}&{\tiny{}0} & {\tiny{}0}&{\tiny{}5822}\tabularnewline
{\tiny{}Traj 7} & {\tiny{}636} & {\tiny{}0}&{\tiny{}0250} & {\tiny{}0}&{\tiny{}0} & {\tiny{}0}&{\tiny{}5030} & {\tiny{}0}&{\tiny{}0100} & {\tiny{}0}&{\tiny{}0} & {\tiny{}0}&{\tiny{}3001} & {\tiny{}0}&{\tiny{}1560} & {\tiny{}0}&{\tiny{}0} & {\tiny{}0}&{\tiny{}4760}\tabularnewline
{\tiny{}Traj 8} & {\tiny{}751} & {\tiny{}0}&{\tiny{}0110} & {\tiny{}0}&{\tiny{}0} & {\tiny{}0}&{\tiny{}3610} & {\tiny{}0}&{\tiny{}0731} & {\tiny{}0}&{\tiny{}0} & {\tiny{}0}&{\tiny{}1935} & {\tiny{}0}&{\tiny{}0200} & {\tiny{}0}&{\tiny{}0} & {\tiny{}0}&{\tiny{}6064}\tabularnewline
{\tiny{}Traj 9} & {\tiny{}650} & {\tiny{}0}&{\tiny{}0030} & {\tiny{}0}&{\tiny{}0} & {\tiny{}0}&{\tiny{}5590} & {\tiny{}0}&{\tiny{}0} & {\tiny{}0}&{\tiny{}0} & {\tiny{}0}&{\tiny{}6010} & {\tiny{}0}&{\tiny{}0060} & {\tiny{}0}&{\tiny{}0} & {\tiny{}0}&{\tiny{}6000}\tabularnewline
{\tiny{}Traj 11} & {\tiny{}375} & {\tiny{}0}&{\tiny{}0070} & {\tiny{}0}&{\tiny{}0} & {\tiny{}0}&{\tiny{}7720} & {\tiny{}0}&{\tiny{}0} & {\tiny{}0}&{\tiny{}0} & {\tiny{}0}&{\tiny{}5679} & {\tiny{}0}&{\tiny{}1220} & {\tiny{}0}&{\tiny{}0} & {\tiny{}0}&{\tiny{}7920}\tabularnewline
\end{tabular}{\tiny\par}

{\tiny{}}%
\begin{tabular}{ccccccccccc}
 & {\tiny{}Steps} & {\tiny{}$\hat{p}_{23}$} &  &  & {\tiny{}$\hat{p}_{31}$} &  &  & {\tiny{}$\hat{p}_{32}$} &  & \tabularnewline
\hline 
{\tiny{}Traj 4} & {\tiny{}448} & {\tiny{}0.0} & {\tiny{}0.0} & {\tiny{}0.8139} & {\tiny{}0.0} & {\tiny{}0.0} & {\tiny{}0.5445} & {\tiny{}0.0093} & {\tiny{}0.0} & {\tiny{}0.8253}\tabularnewline
{\tiny{}Traj 7} & {\tiny{}636} & {\tiny{}0.0} & {\tiny{}0.0} & {\tiny{}0.6980} & {\tiny{}0.0} & {\tiny{}0.0} & {\tiny{}0.2073} & {\tiny{}0.0100} & {\tiny{}0.0} & {\tiny{}0.6520}\tabularnewline
{\tiny{}Traj 8} & {\tiny{}751} & {\tiny{}0.0} & {\tiny{}0.0} & {\tiny{}0.8370} & {\tiny{}0.0} & {\tiny{}0.0} & {\tiny{}0.3889} & {\tiny{}0.0180} & {\tiny{}0.0} & {\tiny{}0.8460}\tabularnewline
{\tiny{}Traj 9} & {\tiny{}650} & {\tiny{}0.0} & {\tiny{}0.0} & {\tiny{}0.6190} & {\tiny{}0.0110} & {\tiny{}0.0} & {\tiny{}0.3110} & {\tiny{}0.0} & {\tiny{}0.0} & {\tiny{}0.7270}\tabularnewline
{\tiny{}Traj 11} & {\tiny{}375} & {\tiny{}0.0050} & {\tiny{}0.0} & {\tiny{}0.7100} & {\tiny{}0.0030} & {\tiny{}0.0} & {\tiny{}0.6560} & {\tiny{}0.0830} & {\tiny{}0.0} & {\tiny{}0.7560}\tabularnewline
\end{tabular}{\tiny{}\label{tab:pdk1_3state_trates}}{\tiny\par}

{\tiny{}\caption[3-State HMM transition probability MAPs and 95\% credible intervals]{MLEs and associated 95\% credible intervals for transition probabilities
from a 3-state model analysis of PDK1 trajectories.}
}{\tiny\par}
\end{table}

\begin{table}[!htb]
\centering
{\tiny{}}%
\begin{tabular}{ccr@{\extracolsep{0pt}.}lr@{\extracolsep{0pt}.}lr@{\extracolsep{0pt}.}lr@{\extracolsep{0pt}.}lr@{\extracolsep{0pt}.}lr@{\extracolsep{0pt}.}lr@{\extracolsep{0pt}.}lr@{\extracolsep{0pt}.}lr@{\extracolsep{0pt}.}l}
 & {\tiny{}Steps} & \multicolumn{2}{c}{{\tiny{}$\hat{k}_{12}$}} & \multicolumn{2}{c}{} & \multicolumn{2}{c}{} & \multicolumn{2}{c}{{\tiny{}$\hat{k}_{13}$}} & \multicolumn{2}{c}{} & \multicolumn{2}{c}{} & \multicolumn{2}{c}{{\tiny{}$\hat{k}_{21}$}} & \multicolumn{2}{c}{} & \multicolumn{2}{c}{}\tabularnewline
\hline 
{\tiny{}Traj 4} & {\tiny{}448} & {\tiny{}2}&{\tiny{}5} & {\tiny{}0}&{\tiny{}0} & {\tiny{}15}&{\tiny{}} & {\tiny{}1}&{\tiny{}0} & {\tiny{}0}&{\tiny{}0} & {\tiny{}8}&{\tiny{}8} & {\tiny{}0}&{\tiny{}6} & {\tiny{}0}&{\tiny{}0} & {\tiny{}29}&{\tiny{}}\tabularnewline
{\tiny{}Traj 7} & {\tiny{}636} & {\tiny{}1}&{\tiny{}3} & {\tiny{}0}&{\tiny{}0} & {\tiny{}25}&{\tiny{}} & {\tiny{}0}&{\tiny{}5} & {\tiny{}0}&{\tiny{}0} & {\tiny{}15}&{\tiny{}0} & {\tiny{}7}&{\tiny{}8} & {\tiny{}0}&{\tiny{}0} & {\tiny{}24}&{\tiny{}}\tabularnewline
{\tiny{}Traj 8} & {\tiny{}751} & {\tiny{}0}&{\tiny{}55} & {\tiny{}0}&{\tiny{}0} & {\tiny{}18}&{\tiny{}} & {\tiny{}3}&{\tiny{}7} & {\tiny{}0}&{\tiny{}0} & {\tiny{}9}&{\tiny{}7} & {\tiny{}1}&{\tiny{}0} & {\tiny{}0}&{\tiny{}0} & {\tiny{}30}&{\tiny{}}\tabularnewline
{\tiny{}Traj 9} & {\tiny{}650} & {\tiny{}0}&{\tiny{}15} & {\tiny{}0}&{\tiny{}0} & {\tiny{}28}&{\tiny{}} & {\tiny{}0}&{\tiny{}0} & {\tiny{}0}&{\tiny{}0} & {\tiny{}30}&{\tiny{}} & {\tiny{}0}&{\tiny{}30} & {\tiny{}0}&{\tiny{}0} & {\tiny{}30}&{\tiny{}}\tabularnewline
{\tiny{}Traj 11} & {\tiny{}375} & {\tiny{}0}&{\tiny{}35} & {\tiny{}0}&{\tiny{}0} & {\tiny{}39}&{\tiny{}} & {\tiny{}0}&{\tiny{}0} & {\tiny{}0}&{\tiny{}0} & {\tiny{}28}&{\tiny{}} & {\tiny{}6}&{\tiny{}1} & {\tiny{}0}&{\tiny{}0} & {\tiny{}39}&{\tiny{}}\tabularnewline
\end{tabular}{\tiny\par}

{\tiny{}}%
\begin{tabular}{r@{\extracolsep{0pt}.}lr@{\extracolsep{0pt}.}lr@{\extracolsep{0pt}.}lr@{\extracolsep{0pt}.}lr@{\extracolsep{0pt}.}lr@{\extracolsep{0pt}.}lr@{\extracolsep{0pt}.}lr@{\extracolsep{0pt}.}lr@{\extracolsep{0pt}.}lr@{\extracolsep{0pt}.}lr@{\extracolsep{0pt}.}l}
\multicolumn{2}{c}{} & \multicolumn{2}{c}{{\tiny{}Steps}} & \multicolumn{2}{c}{{\tiny{}$\hat{k}_{23}$}} & \multicolumn{2}{c}{} & \multicolumn{2}{c}{} & \multicolumn{2}{c}{{\tiny{}$\hat{k}_{31}$}} & \multicolumn{2}{c}{} & \multicolumn{2}{c}{} & \multicolumn{2}{c}{{\tiny{}$\hat{k}_{32}$}} & \multicolumn{2}{c}{} & \multicolumn{2}{c}{}\tabularnewline
\hline 
\multicolumn{2}{c}{{\tiny{}Traj 4}} & \multicolumn{2}{c}{{\tiny{}448}} & {\tiny{}0}&{\tiny{}0} & {\tiny{}0}&{\tiny{}0} & {\tiny{}41}&{\tiny{}} & {\tiny{}0}&{\tiny{}0} & {\tiny{}0}&{\tiny{}0} & {\tiny{}27}&{\tiny{}} & {\tiny{}0}&{\tiny{}47} & {\tiny{}0}&{\tiny{}0} & {\tiny{}41}&{\tiny{}}\tabularnewline
\multicolumn{2}{c}{{\tiny{}Traj 7}} & \multicolumn{2}{c}{{\tiny{}636}} & {\tiny{}0}&{\tiny{}0} & {\tiny{}0}&{\tiny{}0} & {\tiny{}35}&{\tiny{}} & {\tiny{}0}&{\tiny{}0} & {\tiny{}0}&{\tiny{}0} & {\tiny{}10}&{\tiny{}} & {\tiny{}0}&{\tiny{}50} & {\tiny{}0}&{\tiny{}0} & {\tiny{}33}&{\tiny{}}\tabularnewline
\multicolumn{2}{c}{{\tiny{}Traj 8}} & \multicolumn{2}{c}{{\tiny{}751}} & {\tiny{}0}&{\tiny{}0} & {\tiny{}0}&{\tiny{}0} & {\tiny{}42}&{\tiny{}} & {\tiny{}0}&{\tiny{}0} & {\tiny{}0}&{\tiny{}0} & {\tiny{}19}&{\tiny{}} & {\tiny{}0}&{\tiny{}90} & {\tiny{}0}&{\tiny{}0} & {\tiny{}42}&{\tiny{}}\tabularnewline
\multicolumn{2}{c}{{\tiny{}Traj 9}} & \multicolumn{2}{c}{{\tiny{}650}} & {\tiny{}0}&{\tiny{}0} & {\tiny{}0}&{\tiny{}0} & {\tiny{}31}&{\tiny{}} & {\tiny{}0}&{\tiny{}55} & {\tiny{}0}&{\tiny{}0} & {\tiny{}16}&{\tiny{}} & {\tiny{}0}&{\tiny{}0} & {\tiny{}0}&{\tiny{}0} & {\tiny{}36}&{\tiny{}}\tabularnewline
\multicolumn{2}{c}{{\tiny{}Traj 11}} & \multicolumn{2}{c}{{\tiny{}375}} & {\tiny{}0}&{\tiny{}25} & {\tiny{}0}&{\tiny{}0} & {\tiny{}35}&{\tiny{}6} & {\tiny{}0}&{\tiny{}15} & {\tiny{}0}&{\tiny{}0} & {\tiny{}33}&{\tiny{}} & {\tiny{}4}&{\tiny{}2} & {\tiny{}0}&{\tiny{}0} & {\tiny{}38}&{\tiny{}}\tabularnewline
\end{tabular}

{\tiny{}\caption[3-State HMM first-order rate constant MAPs and 95\% credible intervals]{MLEs and associated 95\% credible intervals for first-order rate
constants from a 3-state model analysis of PDK1 trajectories. Rate constants
are reported in units $\text{s}^{-1}.$}\label{tab:pdk1_3state_krates}
}{\tiny\par}
\end{table}

\end{document}